%Paper: hep-th/9205102
%From: ZUCCHINIR@bologna.infn.it
%Date: Thu, 28 MAY 92 09:18 GMT

% This paper consists of two separate plain TeX programs: 1) WTITLE for the
% title page and 2) WTEXT for the main text. Run the programs separately.

%////////////////////////////  File WTITLE.tex  \\\\\\\\\\\\\\\\\\\\\\\\\\\\

\magnification=1200
\baselineskip= .55cm plus .055cm minus .055cm
\vsize=23.8 truecm
\hsize=16.5 truecm
\nopagenumbers

\hbox to 16.5 truecm{May 1992   \hfil DFUB--92}
\hbox to 16.5 truecm{Version 1  \hfil}
\vskip2cm
\centerline{{\bf LIGHT CONE} ${\bf W}_{\bf n}$ {\bf GEOMETRY AND ITS
SYMMETRIES}}
\centerline{\bf AND PROJECTIVE FIELD THEORY}
\vskip1cm
\centerline{by}
\vskip.5cm
\centerline{\bf Roberto Zucchini}
\centerline{\it Dipartimento di Fisica, Universit\`a degli Studi di Bologna}
\centerline{\it V. Irnerio 46, I-40126 Bologna, Italy}
\vskip1cm
\centerline{\bf Abstract}
\vskip.4cm
\noindent
I show that the generalized Beltrami differentials and projective connections
which appear naturally in induced light cone $W_n$ gravity are geometrical
fields parametrizing in one-to-one fashion generalized projective structures
on a fixed base Riemann surface.
I also show that $W_n$ symmetries are nothing but
gauge transformations of the flat ${SL}(n,{\bf C})$ vector bundles canonically
associated to the generalized projective structures. This provides an original
formulation of classical light cone $W_n$ geometry. From the knowledge of the
symmetries, the full BRS algebra is derived. Inspired by the results of
recent literature, I argue that quantum $W_n$ gravity may be formulated as an
induced gauge theory of generalized projective connections. This leads to
projective field theory. The possible anomalies arising at the quantum level
are analyzed by solving Wess-Zumino consistency conditions. The implications
for induced covariant $W_n$ gravity are briefly discussed. The results
presented, valid for arbitrary $n$, reproduce those obtained for $n=2,3$ by
different methods.

\bye

% ///////////////////////////       end        \\\\\\\\\\\\\\\\\\\\\\\\\\\\

% ///////////////////////////  File WTEXT.tex  \\\\\\\\\\\\\\\\\\\\\\\\\\\\\

\magnification=1200
\baselineskip=.55cm plus .55mm minus .55mm
\def\ref#1{\lbrack#1\rbrack}
\font\sf=cmss10
\def\sans#1{\hbox{\sf #1}}

\centerline{\bf 1. Introduction.}
\vskip.5cm

During the last few years, a large body of literature has been devoted to
the study of $W$ algebras and to the understanding of their field
theoretic realizations. Originally introduced as higher spin extensions of
the Virasoro algebra \ref{1--2}, they were later shown to appear naturally
in several contexts, such as cosets of affine Lie algebras \ref{3--4},
gauged WZWN models \ref{5--6}, Toda field theory \ref{7--8}, reductions of
the KP hierarchy \ref{9--12} and, lastly, random matrix models \ref{13--15}.
Coupling the generators of $W_n$ algebras to background fields yields
induced $W_n$ gravity \ref{16-18}. It is a common belief that a satisfactory
formulation of $W_n$ gravity requires a prior understanding of the underlying
geometry and its symmetries. Though progress has been made in this direction
\ref{18--26}, much remains to be understood. In this respect, it is necessary
to distinguish between light cone and covariant $W_n$ geometry. The aim of
this paper is showing how light cone $W_n$ geometry and symmetries thereof
can be understood in terms of deformation theory of generalized projective
structures on a fixed curve, to assess which indications about covariant $W_n$
geometry may result from this geometrical framework and to analyze its field
theoretic implications. I shall now review briefly basic known facts about
$W_n$ gravity.

A number of approaches to light cone $W_n$ geometry have been developed. They
share to various extent the following intuitions. The Virasoro algebra $W_2$
expresses the operator product expansion properties of the energy momentum
tensor $T$. As well-known, $T$ represents the response of a conformal
system to deformations of the underlying conformal geometry parametrized by
the Beltrami differential $\mu$. As $W_n$ algebras are higher spin
extensions of $W_2$ possessing $n-1$ generators, light cone $W_n$ geometry
should be parametrized by $n-1$ generalized Beltrami differentials $\mu_i$
\ref {21--23}. However, the case $n=2$ is exceptional in several ways.
In the general case, because of the non
linearity of the $W_n$ algebra, the induced light cone $W_n$ gravity action
$\Gamma^*_n(\mu)$, obtained by integrating out the matter fields, has an
expansion in inverse powers of the central charge $c$ of the form
$$\Gamma^*_n(\mu)=c\sum_{l=0}^\infty c^{-l}\Gamma_{nl}(\mu)\eqno(1.1)$$
\ref{27--28}. Further, the symmetry transformations of the generalized Beltrami
differentials $\mu_i$ depend on the matter content of the system.
As a consequence, $\Gamma^*_n(\mu)$ obeys a Ward identity in which both the
symmetry transformations of the $\mu_i$'s and the anomaly depend non locally
on the $\mu_i$'s through the functional derivatives of $\Gamma^*_n(\mu)$
\ref{27--28}.

The induced light cone $W_n$ gravity action $\Gamma^*_n(\mu)$ is an effective
action for quantum light cone $W_n$ gravity. The generating functional
$W^*_n(\rho)$ of the connected Green functions of the generalized Beltrami
differentials $\mu_i$ depends on $n-1$ background fields $\rho^i$, which are
generalized projective connections. It has been argued in ref. \ref{27--28}
that $W^*_n(\rho)$ has a very simple structure:
$$W^*_n(\rho)=2k_cW^*_{n0}(Z_c\rho),\eqno(1.2)$$
where $k_c=c/24+O(1)$ and $Z_c=O(1/c)$ is a diagonal matrix and
$W^*_{n0}(\rho)$
is the functional Legendre transform of the leading term $\Gamma^*_{n0}(\mu)$
of the $1/c$ expansion $(1.1)$:
$$W^*_{n0}(\rho)=\Gamma^*_{n0}(\mu)-{1\over\pi}\int_\Sigma d^2z\mu_i\rho^i.
\eqno(1.3)$$
Although the analysis of refs. \ref{27--28} has been carried out in detail only
for $n=2,3$ and to finite order in the $1/c$ expansion, it is believed that
the overall pattern should hold for arbitrary $n$ and non perturbatively in
$1/c$.

Progress has also been made toward a deeper understanding of induced covariant
$W_n$ gravity and its geometry, albeit in this case many problems remain open.
It is by now realized that classical covariant $W_n$ geometry is the extrinsic
geometry of a class of embeddings of the $2d$ base manifold into a $n-1$
dimensional Kaehler manifold \ref{24--26}. The Gauss Codazzi equations of
the embeddings have the form of the $A_n$ Toda equations \ref{26}. A proof
of a holomorphic factorization theorem for $n>2$ on the same lines as that of
refs. \ref{31--32} is yet to be achieved, though there are arguments and
partial calculations of the Quillen-Belavin-Knizhnik anomaly for $n=3$
showing the expected connection with $A_n$ Toda field theory \ref{33-34}.
A clarification of the $W_n$ generalization of ordinary Weyl and
diffeomorphism invariance would be desirable.

The point of view adopted in this paper is the following. The generating
functional $W^*_n(\rho)$ of the connected Green functions of the generalized
Beltrami differentials $\mu_i$ is the main object of interest in quantum
light cone $W_n$ gravity. Up to field renormalizations, $W^*_n(\rho)$ is
proportional to the functional $W^*_{n0}(\rho)$ defined in $(1.3)$.
The analysis of anomalies carried out in refs. \ref{27--30} shows that
$W^*_{n0}(\rho)$ obeys a Ward identity in which the symmetry transformations of
the $\rho^i$'s and the anomaly are given by local expressions in the
$\rho^i$'s themselves and the ghost fields. This suggests that $W^*_{n0}(\rho)$
may be realized as the induced gauge action of an ordinary local gauge theory
whose basic gauge fields are the $\rho^i$'s. Models such as the ones just
described will be called chiral projective field theories for reference and
have already appeared for $n=2,3$ in the analysis of refs. \ref{27--28}.
Their interest is manifold. They provide a novel class of local gauge theories
characterized by classical light cone $W_n$ geometry and may be viewed as
effective realizations of quantum $W_n$ gravity. Quantization the $\rho^i$'s
via the non local induced projective action $\Gamma_n(\rho)$ yields the
generating functional $W_n(\mu)$ of the connected Green functions of the
$\rho^i$'s. Upon rescaling conventionally $\Gamma(\rho)$ by a coefficient
$2k$ working as the loop counting parameter $\hbar^{-1}$, $W_n(\mu)$ has a
$1/k$ expansion analogous to $(1.1)$. Assuming that $k$ is chosen so that the
anomaly of $\Gamma(\rho)$ equals that of $W^*_{n0}(\rho)$, it is conceivable
that, for $k=k_c$, $W_n(Z_c{}^{-1}\mu)$ is the generating functional of a
$W_n$ algebra with central charge $c$, as $k_c\rightarrow c/24$ as
$c\rightarrow\infty$. However, I have no proof of this conjecture.

The purpose of this paper is twofold. First, I will attempt a systematic study
of classical light cone $W_n$ geometry and on a compact
connected surface of high genus. Such choice has been dictated by the desire
to highlight certain geometric features that would be trivialized in the planar
topology. This is also meant to be a development of the approach initiated in
refs. \ref{22--23}. I shall provide a geometrical interpretation of the
generalized Beltrami differentials and projective connections in terms of
deformation theory of generalized projective structures on a curve with a
fixed conformal structure. Developing on the ideas of ref. \ref{26},
I shall also show how the geometric fields of light cone $W_n$
geometry describe deformations of embeddings of the base curve into a target
complex $n-1$ dimensional manifold admitting projective coordinate structures.
In particular, I shall indicate how
the notion of degenerated curve, which arises naturally in ordinary conformal
geometry, may be generalized to $W_n$ geometry. It will emerge from the
discussion how classical light cone $W_n$ geometry extends ordinary $W_2$
geometry and an attempt to make contact with covariant $W_n$
geometry will be made.
I hope this will shed light on the interpretation of the geometric
fields, a problem only partially solved in earlier literature \ref{21--25}.
Second, I shall show that $W_n$ symmetries are just gauge transformations
of the flat $SL(n,{\bf C})$ vector bundle canonically associated to the
underlying generalized projective structure. From the knowledge of the
symmetries, I will
construct the BRS algebra. I shall then move to a study of chiral projective
field theory at the classical and quantum level by solving the Wess-Zumino
consistency condition to find the relevant anomaly and the Ward identity.
\vskip.6cm
\centerline{{\bf 2. Classical Light Cone} ${\bf W}_{\bf n}$ {\bf Geometry.}}
\vskip.5cm

In this section, I shall give an illustration of classical light cone $W_n$
geometry that highlights the way it generalizes $W_2$ geometry.

Consider a compact connected curve $\Sigma$ of genus $g\geq 2$ endowed with a
fixed reference conformal structure $\sans a$. I shall denote the generic
coordinate of $\sans a$ by $z$ and partial differentiation with respect to $z$
by $\partial$ appending early Latin indices $a$, $b$, $c$, ... to distinguish
between different coordinates of $\sans a$ whenever necessary. The
$\sans a$--holomorphic canonical line bundle $k$ is defined as usual by the
transition functions $k_{ba}=\partial_bz_a$ and a fixed choice of the tensor
square root $k^{1\over 2}$ is made \ref{35}.

A projective structure \ref{35} $\sans A$ on $\Sigma$ is a maximal collection
of local maps $Z_\alpha$ of $\Sigma$ into $\bf C$ with the following
properties.
The domains of the $Z_\alpha$'s cover $\Sigma$. On overlapping domains one has
$$Z_\beta={a_{\beta\alpha}Z_\alpha+b_{\beta\alpha}\over
c_{\beta\alpha}Z_\alpha+d_{\beta\alpha}},\eqno(2.1)$$
where $\Big(\matrix{a_{\beta\alpha}&b_{\beta\alpha}\cr
c_{\beta\alpha}&d_{\beta\alpha}\cr}\Big)$ is a nonsingular constant complex
matrix defined up to overall normalization. Further, the following non
singularity condition
$$\partial Z\not=0\quad{\rm pointwise}\eqno(2.2)$$
is fulfilled. It is known that projective structures are parametrized in
one-to-one fashion by a pair of fields $\mu$ and $\rho$ defined by
$$\rho=-(1/2)\{Z,z\}, \eqno(2.3)$$
$$\mu=\bar\partial Z/\partial Z, \eqno(2.4)$$
where $\{f,\zeta\}=-2(\partial_\zeta f)^{1\over 2}\partial^2(\partial_\zeta f)^
{-{1\over 2}}$ is the Schwarzian derivative of $f$ with respect to $\zeta$.
$\mu$ and $\rho$ are independent of the choice of the map  $Z$ in
$\sans A$ and glue under coordinate changes in $\sans a$ as follows:
$$\rho_b=k^2{}_{ba}\big[\rho_a+(1/2)\{z_b,z_a\}\big],\eqno(2.5)$$
$$\mu_b=\bar k_{ba}k^{-1}{}_{ba}\mu_a.\eqno(2.6)$$
They further satisfy the following compatibility condition:
$$\big(\bar\partial-\mu\partial-2(\partial\mu)\big)\rho=-(1/2)\partial^3\mu.
\eqno(2.7)$$
In the standard definition of projective structure, the non singularity
condition $\partial Z\bar\partial\bar Z-\partial\bar Z\bar\partial Z>0$ is
imposed instead of $(2.2)$. Such condition is stronger than $(2.2)$,
implying besides $(2.2)$ the bound $|\mu|<1$.
The formulation I am adopting here is more general than the customary one.
Later, I shall show how to revert to the more restrictive formulation in a
way indicating a generalization to $W_n$ geometry.

Classical light cone $W_n$ geometry is based on a rather straightforward
generalization of the notion of projective structure. A generalized
projective structure is a maximal collection of local maps $Z_\alpha$
$=$ $(Z_\alpha{}^1,\cdots,Z_\alpha{}^{n-1})$ of $\Sigma$
into ${\bf C}^{n-1}$ with the following properties. The domains of
the $Z_\alpha$'s cover $\Sigma$. On overlapping domains one has that
$$Z_\beta{}^r={\tilde\Phi_{\beta\alpha}{}^r{}_sZ_\alpha{}^s\over
\tilde\Phi_{\beta\alpha}{}^0{}_tZ_\alpha{}^t},\eqno(2.8)$$
where $\tilde\Phi_{\beta\alpha}$ is a constant non singular $n\times n$
complex matrix defined up to overall normalization, sum over repeated indices
is implied and $Z_\alpha{}^0=1$ by convention. Further, defining
$$\tilde W_i{}^r=\partial^iZ^r,\quad i,r=0,\cdots,n-1,\eqno(2.9)$$
the non singularity condition
$$\Delta={\rm det}\tilde W\not=0\quad{\rm pointwise}\eqno(2.10)$$
holds. As is straightforward to verify, such condition is compatible with
map changes in $\sans A$ and coordinate changes in $\sans a$. Now, the natural
question arises on whether the family of generalized projective structures
can be parametrized by a set of $2n-2$ geometrical fields $\mu_i$ and
$\rho^i$ analogous to $\mu$ and $\rho$ (cfr. eqs. $(2.3)-(2.4)$) and satisfying
some compatibility condition (cfr. eq. $(2.7)$. The answer
is affirmative. What is more is that such fields are precisely the geometrical
fields entering in $W_n$ gravity as illustrated in the introduction.
The rest of the section will be dedicated to the construction of these fields
and to the study of their properties.

By $(2.9)$, the field $\Delta$ is nowhere vanishing. So, for any given
coordinate $z$ in $\sans a$ and any map $Z$ in $\sans A$, one can choose a
smooth branch $\Delta^{1\over n}$ of the $n$--th root of $\Delta$ free of
branch points. The choice of the branch can be made in such a way that on
overlapping domains
$$\Delta^{1\over n}{}_{b\beta}{}={c_{\beta\alpha}k^{{1\over
2}(n-1)}{}_{ba}\over
\tilde\Phi_{\beta\alpha}{}^0{}_sZ_\alpha{}^s}\Delta^{1\over n}{}_{a\alpha},
\eqno(2.11)$$
where $c_{\beta\alpha}$ is a non-zero constant independent from coordinate
choices in $\sans a$ and such that
$$c_{\beta\alpha}{}^n={\rm det}\tilde\Phi_{\beta\alpha}.\eqno(2.12)$$
Setting
$$\Phi_{\beta\alpha}=c_{\beta\alpha}{}^{-1}\tilde\Phi_{\beta\alpha},
\eqno(2.13)$$
one has that
$${\rm det}\Phi_{\beta\alpha}=1, \eqno(2.14)$$
$$\Phi_{\alpha\alpha}=1,\eqno(2.15a)$$
$$\Phi_{\gamma\beta}\Phi_{\beta\alpha}\Phi_{\alpha\gamma}=1.\eqno(2.15b)$$
So, $\Phi$ is a $SL(n,{\bf C})$--valued 1--cocycle defining a flat
$SL(n,{\bf C})$ vector bundle on $\Sigma$ \ref{36}. Such bundle is canonically
associated to the generalized projective structure $\sans A$.

\noindent
{\it Proof}. From $(2.8)$, it follows readily that
$$\tilde\Phi_{\alpha\alpha}=\lambda_\alpha1,\eqno(2.16a)$$
$$\tilde\Phi_{\gamma\beta}\tilde\Phi_{\beta\alpha}\tilde\Phi_{\alpha\gamma}=
\kappa_{\gamma\beta\alpha}1,\eqno(2.16b)$$
on overlapping domains, where $\lambda_\alpha$ and $\kappa_{\gamma\beta\alpha}$
are non-zero constants. It is straightforward to check that $(2.11)-(2.12)$
hold in general with $c_{\beta\alpha}$ replaced by a non-zero constant
$c_{b\beta a\alpha}$ and that
$${c_{b\beta a\alpha}\over\tilde\Phi_{\beta\alpha}{}^0{}_sZ_\alpha{}^s}=
{c_{b\beta c\gamma}c_{c\gamma d\delta}c_{d\delta a\alpha}\over
\tilde\Phi_{\beta\gamma}\tilde\Phi_{\gamma\delta}
\tilde\Phi_{\delta\alpha}{}^0{}_sZ_\alpha{}^s}.\eqno(2.17)$$
For $b\beta=a\alpha$ the left hand side of $(2.17)$ equals 1. By using
this fact and $(2.16a)$, it is straightforward to check that
$\eta^{(\alpha)}_{ba}=c_{b\alpha a\alpha}/\lambda_\alpha$ is a
${\bf Z}_n$--valued 1--cocycle on the domain of $Z_\alpha$. Since the latter
can be assumed to be simply connected, this 1--cocycle is a 1--coboundary.
Hence, $\eta^{(\alpha)}_{ba}=\eta^{(\alpha)}_b\eta^{(\alpha)}_{\vphantom{b}a}{}
^{-1}$,
where $\eta^{(\alpha)}_c$ is a ${\bf Z}_n$--valued 0--cocycle on the domain of
$Z_\alpha$. By redefining the branch of $\Delta^{1\over n}{}_{a\alpha}$ into
$\eta^{(\alpha)}_a{}^{-1}\Delta^{1\over n}{}_{a\alpha}$, one can assume that
$c_{b\alpha a\alpha}=\lambda_\alpha$, i. e. $c_{b\alpha a\alpha}$ is
independent from any coordinate choice in $\sans a$. By using this fact,
$(2.16a)$ and $(2.17)$ with $\gamma=\beta$ and $\delta=\alpha$, it follows
readily that there is a constant $c_{\beta\alpha}$ independent from any
coordinate choice in $\sans a$ and such that $c_{b\beta a\alpha}=
c_{\beta\alpha}$. By using $(2.16)-(2.17)$, it is immediately checked that
$$c_{\alpha\alpha}=\lambda_\alpha,\eqno(2.18a)$$
$$c_{\gamma\beta}c_{\beta\alpha}c_{\alpha\gamma}=\kappa_{\gamma\beta\alpha},
\eqno(2.18b)$$
from which $(2.14)-(2.15)$ follow easily. {\it QED}.

Next, define
$$W_i{}^r=\partial^i\big(\Delta^{-{1\over n}}Z^r\big),\quad i,r=0,\cdots,n-1.
\eqno(2.19)$$
A straightforward analysis shows that
$${\rm det}W=1\eqno(2.20)$$
identically and that, under coordinate changes in $\sans a$ and map
changes in $\sans A$,
$$W_{b\beta}=\Lambda_{ba}W_{a\alpha}\Phi^\vee{}_{\alpha\beta},\eqno(2.21)$$
where $\Lambda$ is the bundle of $n-1$--jets of
sections of $k^{{1\over 2}(1-n)}$ and $\Phi^\vee$ is the dual of the bundle
$\Phi$. The jet bundle $\Lambda$ is defined as follows. If $\psi$ is a smooth
section of $k^{{1\over 2}(1-n)}$, the $n-1$--jet of $\psi$ is by definition
the $n$--tuple of fields
$$j_{n-1}\psi=\big(\psi,\partial\psi,\cdots,\partial^{n-1}\psi\big)^t.
\eqno(2.22)$$
The jet bundle $\Lambda$ describes the gluing of the jet $j_{n-1}\psi$
under coordinate changes in $\sans a$:
$$j_{n-1}\psi_b=\Lambda_{ba}j_{n-1}\psi_a.\eqno(2.23)$$
Hence, $\Lambda$ is a rank $n$ $\sans a$--holomorphic vector bundle. The
transition matrices of $\Lambda$ can be computed from the formula
$$\Lambda_{ba}{}_i{}^j=\sum_{w\in {\cal W}_{i-j,j}}w(k_{ba}\partial_a, k_{ba})
k^{{1\over 2}(1-n)}{}_{ba},\eqno(2.24)$$
where ${\cal W}_{i,j}$ is the set of all words $w$ of two letters $x$ and $y$
containing $i$ occurrences of $x$ and $j$ occurrences of $y$ and, for any two
operators $X$ and $Y$ and any word $w$, $w(X,Y)$ is the operator obtained by
replacing all occurrences of $x$ and $y$ in $w$ by $X$ and $Y$, respectively.
{}From $(2.24)$, it can be checked immediately that all transition matrices of
$\Lambda$ are lower triangular and have unit determinant.
These are indeed the only properties of $\Lambda$ that will be
needed in the following discussion. The dual bundle $\Phi^\vee$ of $\Phi$ is
defined by $\Phi^\vee{}_{\alpha\beta}=\Phi_{\alpha\beta}{}^{-1t}$.
Eq. $(2.21)$ states that the jet bundle $\Lambda$ and the flat $SL(n,{\bf C})$
vector bundle $\Phi$ are smoothly equivalent. In other words, $\Lambda$,
viewed as a smooth bundle, admits a flat form $\Phi$ \ref{36}. The matrix $W$
precisely represents the local change of frame leading from the lower
triangular to the flat form of the bundle.

\noindent{\it Proof}. The proof of $(2.20)$ follows from $(2.10)$, $(2.19)$
and the Leibniz formula. $(2.21)$ follows from $(2.8)$ and $(2.11)$ with
$\tilde\Phi$ replaced by $\Phi$. {\it QED}.

{}From the matrix $(2.25)$ one can define the following fields:
$$\Omega=\partial W W^{-1},\eqno(2.25a)$$
$$\Omega^*=\bar\partial W W^{-1}.\eqno(2.25b)$$
The salient property of $\Omega$ and $\Omega^*$ is that they are independent
from map choices in $\sans A$. One also easily verifies that
$$\Omega_b=k_{ba}\big[\Lambda_{ba}\Omega_a\Lambda_{ba}{}^{-1}
+\partial_a\Lambda_{ba}\Lambda_{ba}{}^{-1}\big],\eqno(2.26a)$$
$$\Omega^*{}_b=\bar k_{ba}\Lambda_{ba}\Omega^*{}_a\Lambda_{ba}{}^{-1},
\eqno(2.26b)$$
under coordinate changes in $\sans a$. Further,
$${\rm tr}\Omega=0,\eqno(2.27a)$$
$${\rm tr}\Omega^*=0,\eqno(2.27b)$$
$$\bar\partial\Omega-\partial\Omega^*+[\Omega,\Omega^*]=0.\eqno(2.28)$$
$(2.26)-(2.28)$ show that $\Omega$ and $\Omega^*$ are the two
components of a flat connection of the jet bundle $\Lambda$.

Not all matrix elements of $\Omega$ and $\Omega^*$ are independent.
They can be expressed in terms of the following independent elements:
$$\rho^i=\Omega_{n-1}^{\hphantom{1}i},\quad i\leq n-2,\eqno(2.29)$$
$$\mu_j=\Omega^*{}_j{}^{n-1},\quad j\leq n-2.\eqno(2.30)$$
The explicit expressions of all elements are
$$\eqalignno{\Omega_i{}^j&=\delta_{i+1,j},\quad i\leq n-2,&(2.31a)\cr
\Omega_{n-1}{}^j&=\rho^j,\quad j\leq n-2,&(2.31b)\cr
\Omega_{n-1}^{\hphantom{1}n-1}&=0,&(2.31c)\cr}$$
$$\eqalignno{\Omega^*{}_i{}^j&=-\sum_{k=1}^{n-2-j}\sum_{l=0}^{n-2-j-k}(-1)^l
{n-2-j-k\choose l}\partial^l\big(\mu_{i+n-2-j-k-l}\rho^{n-1-k}\big)&(2.32a)\cr
&\hskip2.7cm+\sum_{l=0}^{n-1-j}(-1)^l{n-1-j\choose l}\partial^l\mu_{i+n-1-j-l}
,\quad i<j,&\cr
\Omega^*{}_i{}^j&=-\sum_{k=1}^{n-2-i}\sum_{l=0}^{n-2-i-k}(-1)^l{n-2-i-k\choose
l}\partial^l\big(\mu_{n-2-k-l}\rho^{n-1-k}\big)&(2.32b)\cr
+&\sum_{l=1}^{n-1-i}(-1)^l{n-1-i\choose l}
\partial^l\mu_{n-1-l}
-{1\over n}\Bigg[\sum_{l=1}^{n-1}(-1)^l{n\choose l+1}
\partial^l\mu_{n-1-l}&\cr
&\hskip.6cm-\sum_{k=1}^{n-2}\sum_{l=0}^{n-2-k}(-1)^l
{n-1-k\choose l+1}\partial^l\big(\mu_{n-2-k-l}\rho^{n-1-k}\big)
\Bigg],\quad i=j,&\cr
\Omega^*{}_i{}^j&=-\sum_{k=1}^{n-2}\sum_{l=0}^{n-2-k}(-1)^l\Bigg[
\sum_{m=0}^{\langle n-2-k-l,j\rangle}{i-1-m\choose i-j-1}{n-2-k-m\choose l}
&(2.32c)\cr
&\hskip1cm-{1\over n}{i\choose j}{n-1-k\choose l+1}\Bigg]\partial^{i-j+l}\big(
\mu_{n-2-k-l}\rho^{n-1-k}\big)&\cr
&\hskip1cm+\sum_{k=0}^j\sum_{l=0}^{i-j-1}{l+k \choose k}
\partial^l\big(\mu_{i-1-k-l}\rho^{j-k}\big)&\cr
+&\sum_{l=1}^{n-1}(-1)^l\Bigg[
\sum_{m=0}^{\langle n-1-l,j\rangle}{i-1-m\choose i-j-1}{n-1-m\choose l}
&\cr
&\hskip3.41cm-{1\over n}{i\choose j}{n\choose l+1}\Bigg]\partial^{i-j+l}
\mu_{n-1-l},\quad i>j, &\cr}$$
where $\langle a,b\rangle={\rm min}\{a,b\}$ and $\sum_{l=p}^qf_l=0$
whenever $q<p$ by convention.

\noindent{\it Proof}. From $(2.19)$, it follows readily that, for $i\leq n-2$,
$\partial W_i{}^r$ $=$ $W_{i+1}^{\hphantom{1}r}$. From this fact and $(2.25a)$,
$(2.31a)$ follows readily. $(2.31b)$ is obvious. $(2.31c)$ follows from
$(2.31a)$ and the tracelessness property $(2.27a)$. By using repeatedly the
relations $\partial W_i{}^r=\Omega_i{}^jW_j{}^r$, $\bar\partial W_i{}^r=
\Omega^*{}_i{}^jW_j{}^r$ and $\bar\partial W_i{}^r=\partial(\bar\partial
W_{i-1}^{\hphantom{1}r})$ with $i\geq 1$, which follow trivially from $(2.19)$
and $(2.25)$, one finds the relations
$$\Omega^*{}_i{}^j-\Omega^*{}_{i-1}^{\hphantom{1}j-1}
-\partial\Omega^*{}_{i-1}^{\hphantom{1}j}=
\Omega^*{}_{i-1}^{\hphantom{1}n-1}\Omega_{n-1}^{\hphantom{1}j},\quad 1\leq i,
\eqno(2.33)$$
where $\Omega^*{}_i^{-1}=0$ by convention.
By setting $i=j$ in $(2.33)$ and summing over $i$ from $k+1$ to $n-1$ with
$0\leq k\leq n-2$, one gets
$$\Omega^*{}_k{}^k=\Omega^*{}_{n-1}^{\hphantom{1}n-1}
-\sum_{i=k+1}^{n-1}\big(\partial\Omega^*{}_{i-1}^{\hphantom{1}i}+
\Omega^*{}_{i-1}^{\hphantom{1}n-1}\Omega_{n-1}^{\hphantom{1}i}\big)\eqno(2.34)$$
(no sum on $k$). By using this identity and the tracelessness relation
$(2.27b)$, one obtains
$$\Omega^*{}_{n-1}^{\hphantom{1}n-1}={1\over n}
\sum_{i=1}^{n-1}i\big(\partial\Omega^*{}_{i-1}^{\hphantom{1}i}+
\Omega^*{}_{i-1}^{\hphantom{1}n-1}\Omega_{n-1}^{\hphantom{1}i}\big).
\eqno(2.35)$$
$(2.33)$ and $(2.35)$ constitute a set of recurrence
relations and a constraint on the matrix elements of $\Omega^*$, respectively.
Let us show that these have a unique solution once
$\Omega^*{}_i^{\hphantom{1}n-1}$ and $\Omega_{n-1}^{\hphantom{1}i}$ are
specified for $i\leq n-2$. First, I shall show uniqueness. Denote by $\omega^*$
the difference of two solutions. Now, $\omega^*{}_i{}^{n-1}=0$ for $i\leq n-2$
since the value of the $(i,n-1)$ matrix element of both solutions of $(2.34)$
and $(2.35)$ is assigned by hypothesis. From $(2.33)-(2.35)$, it follows
then that $\omega^*$ satisfies
$$\omega^*{}_i{}^j-\omega^*{}_{i-1}^{\hphantom{1}j-1}
-\partial\omega^*{}_{i-1}^{\hphantom{1}j}=0,\quad 1\leq i,\eqno(2.36)$$
where $\omega^*{}_i^{-1}=0$ by convention, and
$$\omega^*{}_k{}^k=\omega^*{}_{n-1}^{\hphantom{1}n-1}
-\sum_{i=k+1}^{n-1}\partial\omega^*{}_{i-1}^{\hphantom{1}i},\eqno(2.37)$$
$$\omega^*{}_{n-1}^{\hphantom{1}n-1}={1\over n}
\sum_{i=1}^{n-1}i\partial\omega^*{}_{i-1}^{\hphantom{1}i}.\eqno(2.38)$$
Using that $\omega^*{}_i{}^{n-1}=0$ for $i\leq n-2$, one can show from $(2.36)$
by induction that $\omega^*{}_i{}^j=0$ for $i<j$. From $(2.37)-(2.38)$, it
follows immediately that $\omega^*{}_i{}^j=0$ for $i=j$ too. From this fact and
$(2.36)$ one verifies easily that $\omega^*{}_i{}^j=0$ for $0=j<i$. By
proceeding again by induction, one can show from $(2.36)$ that
$\omega^*{}_i{}^j=0$ for $j<i$. This shows uniqueness. It is now
straightforward albeit lengthy to verify that the expression $(2.32)$
solves $(2.33)$ and $(2.35)$. Since the solution is unique $(2.32)$ provides
the expression of the matrix elements of $\Omega^*$. {\it QED}.

It is clear that the $\rho^i$'s and the $\mu_i$'s are independent from map
choices in $\sans A$. From $(2.26)$, one can also check easily that they
transform as
$$\eqalignno{\rho_b{}^i&=k^{{1\over 2}(1+n)}{}_{ba}\sum_{l=0}^{n-2}
\Lambda_{ba}{}^{-1}{}_l{}^i\rho_a{}^l+\sigma_{ba}{}^i,&(2.39)\cr
\sigma_{ba}{}^i&=k_{ba}\bigg[\partial_a\Lambda_{ba}{}_{n-1}^{\hphantom{1}k}
\Lambda_{ba}{}^{-1}{}_k{}^i+\sum_{j=0}^{n-2}\Lambda_{ba}{}_{n-1}^{\hphantom{1}j}
\Lambda_{ba}{}^{-1}{}_{j+1}^{\hphantom{1}i}\bigg],\cr}$$
$$\mu_{bi}=\bar k_{ba} k^{{1\over 2}(1-n)}{}_{ba}\sum_{l=0}^{n-2}
\Lambda_{ba}{}_i{}^l\mu_{al}\eqno(2.40)$$
under coordinate changes in $\sans a$.
{}From the flatness relation $(2.28)$, it follows further that the $\rho^i$'s
and
the $\mu_i$'s satisfy the compatibility condition
$$\bar\partial\rho^i+\sum_{k=0}^{n-2}\Omega^*{}_k{}^i\rho^k-
\Omega^*{}_{n-1}^{\hphantom{1}n-1}\rho^i-\partial\Omega^*{}_{n-1}^{\hphantom{1}
i}-\Omega^*{}_{n-1}^{\hphantom{1}i-1}=0,\quad i=0,\cdots,n-2,\eqno(2.41)$$
where $\Omega^*{}_{n-1}^{\hphantom{1}-1}=0$ by convention and the matrix
elements $\Omega^*{}_i{}^j$ are given in terms of the $\rho^i$'s and the
$\mu_i$'s by $(2.32)$.

\noindent{\it Proof}. $(2.39)-(2.40)$ follow easily from $(2.26)$ upon taking
$(2.29)-(2.30)$ into account. $(2.28)$ implies a host of relations between
the matrix elements of $\Omega$ and $\Omega^*$. However, as is easy to see,
the tracelessness of $\Omega^*$, $(2.31)$ and the identities arising from the
$(i,j)$ matrix element of $(2.28)$ with $i\leq n-2$ reduce to the relations
$(2.33)$ and $(2.35)$ and thus entail no new relations between the $\rho^i$'s
and $\mu_i$'s. The identity arising from the $(n-1,n-1)$ matrix element of
$(2.28)$ follows from the identity already considered and the tracelessness of
$\Omega$ and $\Omega^*$ and thus add nothing new. So, only the identities
arising from the $(n-1,i)$ matrix elements of $(2.28)$ with $i\leq n-2$
entail new relations between the $\rho^i$'s and $\mu_i$'s. By using $(2.31)
-(2.32)$, such identities can be cast as in $(2.41)$. {\it QED}.

I have thus shown that to any generalized projective structure
$\sans A$ there are associated canonically $2n-2$ geometrical fields
$\rho^i$'s and the $\mu_i$'s independent from map choices in $\sans A$
and satisfying $(2.39)-(2.41)$. The converse is also true, as I shall
show next.

Let $2n-2$ fields $\rho^i$'s and the $\mu_i$'s be given which satisfy $(2.39)-
(2.41)$. By using $(2.31)$ and $(2.32)$ as a definition, on can construct by
means of the $\rho^i$'s and the $\mu_i$'s two matrix fields $\Omega$ and
$\Omega^*$ which, as it is easy to see, satisfy $(2.26)-(2.28)$. One then looks
for $n$-tuples $\psi=(\psi^0,\cdots,\psi^{n-1})^t$ of local sections of
$k^{{1\over 2}(1-n)}$ satisfying the system of equations
$$\big(\partial-\Omega)U=0,\eqno(2.42a)$$
$$\big(\bar\partial-\Omega^*)U=0,\eqno(2.42b)$$
with the constraint
$${\rm det}U=1,\quad{\rm pointwise},\eqno(2.43)$$
where
$$U_i{}^r=\partial^i\psi^r,\quad i,r=0,\cdots,n-1.\eqno(2.44)$$
Such system admits solutions. Under coordinate changes in $\sans a$ and
changes of local solutions, $U$ transforms as in $(2.21)$ with $W$ replaced
by $U$ for some flat $SL(n,{\bf C})$ vector bundle $\Phi$.

\noindent{\it Proof}. Eq. $(2.42a)$ is equivalent to a set of $n$ $n$--th order
differential equations of the form $\sum_{l=0}^na_l\partial^l\psi^r=0$ with
$a_{n-1}=0$. As well-known, such system admits locally $n$ linearly
independent solutions $\psi^0,\cdots,\psi^{n-1}$ which can be normalized so
that
$(2.43)$ holds. The solutions are determined up to linear combinations with
$\sans a$--holomorphic $SL(n,{\bf C})$--valued functions. The non uniqueness is
reduced by imposing that the $\psi^r$'s satisfy $(2.42b)$. The compatibility of
$(2.42a)$ and $(2.42b)$ is ensured by the flatness condition $(2.28)$. The
solution of $(2.42a)-(2.42b)$ is determined up to linear combinations with
constant $SL(n,{\bf C})$ matrices. Consistence with coordinate changes in
$\sans a$ requires that the $\psi^r$'s are local sections of
$k^{{1\over 2}(1-n)}$. By the last two remarks, it is then clear that $U$ obeys
$(2.21)$ for some flat $SL(n,{\bf C})$ vector bundle $\Phi$. {\it QED}

On account of $(2.43)$, one can assume that $\psi^0$ is nowhere vanishing
on its domain of definition without any loss of generality. So the maps
$$Z^r=\psi^r/\psi^0\eqno(2.45)$$
are well defined. A straightforward calculation shows that they satisfy
$(2.8)-(2.10)$, $\Delta$ being $(\psi^0)^{-n}$. Thus, the $Z^r$'s define
a generalized projective structure $\sans A$ canonically associated to the
$\rho^i$'s and the $\mu_i$'s. It should be fairly clear that the $2n-2$
geometric fields associated to $\sans A$ by $(2.25a)-(2.25b)$ are precisely
the $\rho^i$'s and the $\mu_i$'s.

To summarize, I have shown that there is a one-to-one correspondence between
sets of $2n-2$ fields $\rho^i$'s and the $\mu_i$'s satisfying $(2.39)-(2.41)$
and generalized projective structures $\sans A$. The relevance of the
jet bundle $\Lambda$ and of the compatibility condition $(2.41)$ in light
cone $W_n$ geometry has been noticed by several authors earlier \ref{21--26}.
Here, I have indicated a novel formulation in which such structures emerge
rather naturally. As can easily be checked, for $n=2$, $(2.29)-(2.30)$ reduce
into $(2.3)-(2.4)$ and, similarly, $(2.39)-(2.41)$ reduce into $(2.5)-(2.7)$.
So, the construction provided effectively generalizes the customary $n=2$
geometric framework. For $n=3$, the geometric fields that are usually
employed in the literature are $\tilde\rho^0=\rho^0-(1/2)\partial\rho^1$,
$\tilde\rho^1=\rho^1$, $\tilde\mu_0=\mu_0$ and $\tilde\mu_1=\mu_1-(1/2)
\partial\mu_0$ \ref{27--28}.
Such combinations are convenient, since $\tilde\rho^1$ is an
ordinary projective connection (cfr. eq. $(2.5)$) while $\tilde\rho^0$,
$\tilde\mu_0$ and $\tilde\mu_1$ are sections of $k^3$, $\bar k k^{-2}$ and
$\bar k k^{-1}$, respectively. Further, $\tilde\mu_0$ and  $\tilde\mu_1$ are
the background fields coupled to the standard generators of the $W_3$ algebra.
For general $n$, it is shown in ref. \ref{12} that one can express the
$\rho^i$'s in terms of fields $\tilde\rho^i$ and derivatives thereof, such
that $\tilde\rho^{n-2}$ is a projective connection and $\tilde\rho^i$ with
$i<n-2$ is a section of $k^{n-i}$.

Now, I shall try to show how the above framework describes the extrinsic
geometry of a class of embeddings of $\Sigma$ into a higher dimensional
complex manifold. Consider the complex unit ball
$B_{n-1}({\bf C})=\{\zeta|\zeta\in{\bf C}^{n-1},|\zeta|<1\}$. The group
${\rm Aut}(B_{n-1}({\bf C}))$ of complex analytic automorphisms of
$B_{n-1}({\bf C})$ consists of all linear fractional transformations of the
form
$$T^r(\zeta)={T^r{}_s\zeta^s\over T^0{}_s\zeta^s},\eqno(2.46)$$
where $T$ is a constant matrix of $SU(1,n-1)$ and $\zeta^0=1$ by convention
\ref{37}. Let $G$ be a properly discontinuous fixed point free subgroup of
${\rm Aut}(B_{n-1}({\bf C}))$ generated by $2g$ elements $a_1,\cdots,a_g$,
$b_1,\cdots,b_g$ with the relation
$$\prod_{i=1}^ga_ib_ia_i{}^{-1}b_i{}^{-1}=1.\eqno(2.47)$$
Note that $G$ defines an injective group homomorphism of the first homotopy
group $\pi_1(\Sigma)$ of $\Sigma$ into ${\rm Aut}(B_{n-1}({\bf C}))$ \ref{38}.
The topological quotient
$${\bf W}_n=B_{n-1}({\bf C})/G\eqno(2.48)$$
is then a complex manifold of complex dimension $n-1$ \ref{37}.
By $(2.47)-(2.48)$, the topological type of ${\bf W}_n$ depends only on $n$
and the genus $g$ of $\Sigma$. Further, one has that:
$$\pi_1({\bf W}_n)=\pi_1(\Sigma),\eqno(2.49)$$
The complex structure of ${\bf W}_n$, instead,
depends also on the conjugation class of $G$ in
${\rm Aut}(B_{n-1}({\bf C}))$. As a complex manifold, ${\bf W}_n$ admits
projective coordinate structures, that is coordinate coverings
$\{\zeta_\alpha\}$ subordinated to the complex structure whose coordinate
transformations are of the form
$$\zeta_\beta{}^r={T_{\beta\alpha}{}^r{}_s\zeta_\alpha{}^s\over
T_{\beta\alpha}{}^0{}_s\zeta_\alpha{}^s},\eqno(2.50)$$
where $T_{\beta\alpha}$ is a non singular $n\times n$ complex matrix and
$\zeta_\alpha{}^0=1$ by convention. The existence of projective structures
puts further constraints on the topology of ${\bf W}_n$. Indeed, the Chern
classes of ${\bf W}_n$ satisfy the relations
$$c_\nu({\bf W}_n)={1\over n^\nu}{n\choose \nu}c_1({\bf W}_n)^{\wedge\nu},
\quad \nu=1,\cdots, n-1\eqno(2.51)$$
\ref{39}. The above construction is standard in the case $n=2$.
In that instance, $G$ is a Fuchsian group of the first kind and one has simply
$${\bf W}_2=\Sigma\eqno(2.52)$$
\ref{38}. For general $n$, the classification of the subgroups $G$ of
${\rm Aut}(B_{n-1}({\bf C}))$ with the properties stated above is not known
to the best of my knowledge.

Let $i_n:\Sigma\rightarrow{\bf W}_n$ be a regular embedding. That is,  $i_n$
is one-to-one and the differential of $i_n$ has real rank $2$ everywhere in
$\Sigma$. Given a projective structure $\sans P$ on ${\bf W}_n$, one can set
$$Z_\alpha{}^r=\zeta_\alpha^r\circ i_n.\eqno(2.53)$$
The $Z^r$'s are a collection of local complex maps in $\Sigma$ which
automatically satisfy $(2.8)$ on account of $(2.50)$.
The regular embedding $i_n$ is said to be non singular with respect
to $\sans P$ if $(2.10)$ is satisfied. If this is the case, the $Z^r$'s
define a generalized projective structure on $\sans A$. Two regular embeddings
$i_n$ and $i'_n$ which are non singular with respect to the projective
structures ${\sans P}$ and ${\sans P}'$ of ${\bf W}_n$, respectively, are said
equivalent if they yield the same generalized projective structure $\sans A$.
Therefore, the generalized projective connections and the Beltrami
differentials may be viewed as moduli for equivalence classes of regular non
singular embeddings of $\Sigma$ into ${\bf W}_n$. For applications, one
would like to find conditions under which a regular embedding $i_n$ is non
singular for a large class of projective structures $\sans P$.
The situation is simpler for $n=2$.
In that case, a regular orientation preserving embedding $i_2$ is such that
$\partial Z^1\bar\partial\bar Z^1-\bar\partial Z^1\bar\partial Z^1>0$.
This relation is equivalent to $\partial Z^1\not=0$ and $|\mu_0|<1$. Thus,
the embedding is non singular and, what is more, it automatically satisfies
the customary bound on the Beltrami differential. For $n>2$, the regularity
of $i_n$ does not imply its non singularity and the restrictions on the
generalized projective connections and Beltrami differentials implied by
regularity cannot be expressed in simple form. This is an issue that clearly
calls for further investigation. Anyhow, it should be clear
that regularity generalizes the notion of non degeneracy of a curve in the
present context.

The interpretation of classical $W_n$ geometry in terms of extrinsic geometry
has been invoked by several authors in the literature \ref{24--26}. The above
formulation has been inspired by such attempts but it differs essentially from
them, since singular embeddings are excluded and the topology of
the target space ${\bf W}_n$ depends on the genus $g$ of $\Sigma$.
\vskip.6cm
\centerline{{\bf 3. The Symmetries of Classical Light Cone} ${\bf W}_{\bf n}$
{\bf Geometry.}}
\vskip.5cm

In this section, I shall show how the basic symmetries of classical light cone
$W_n$ geometry emerge.

For pedagogical reasons, consider first the case $n=2$. Let $\sans A$ be a
projective structure and let $\rho$ and $\mu$ be the associated projective
connection and Beltrami differential (cfr. eqs. $(2.1)-(2.7)$). To $\sans A$
there si associated a flat $SL(2,{\bf C})$ vector bundle $\Phi$ which is
defined by the matrices $\Big(\matrix{a_{\beta\alpha}&b_{\beta\alpha}\cr
c_{\beta\alpha}&d_{\beta\alpha}\cr}\Big)$ appearing in $(2.1)$ with a suitable
choice of the overall normalization. The bundle $\Phi$ can be thought of as
a functional of $\rho$ and $\mu$. Such map is many-to-one. The natural
question arises about which variation of $\rho$ and $\mu$ do not change the
bundle $\Phi$. In essence, this is all what $W_2$ symmetry is about.
Let us analyze this issue in greater detail. Let $Z$ be a map of the
projective structure $\sans A$. We can deform $Z$ by means of a linear
fractional transformation of the form
$$Z'={pZ+q\over rZ+s},\eqno(3.1)$$
where $\Big(\matrix{p&q\cr r&s\cr}\Big)$ is a smooth
$SL(2,{\bf C})$--valued function. The maps $Z'$ will glue as in
$(2.1)$ with the same $SL(2,{\bf C})$ transition matrix provided
$$\pmatrix{p_\beta&q_\beta\cr r_\beta&s_\beta\cr}
\pmatrix{a_{\beta\alpha}&b_{\beta\alpha}\cr
c_{\beta\alpha}&d_{\beta\alpha}\cr}=
\pmatrix{a_{\beta\alpha}&b_{\beta\alpha}\cr
c_{\beta\alpha}&d_{\beta\alpha}\cr}
\pmatrix{p_\alpha&q_\alpha\cr r_\alpha&s_\alpha\cr}.\eqno(3.2)$$
Geometrically, this means that the matrix functions $\Xi=\Big(\matrix{p&q\cr
r&s\cr}\Big)$ define a finite gauge transformation of the flat $SL(2,{\bf C})$
vector bundle $\Phi$. However, it is very difficult to ascertain which gauge
transformations $\Xi$ are such that the maps $Z'$ satisfy the non
singularity condition $(2.2)$. For such reason, it is sounder to switch to
infinitesimal version of the above construction. One consider infinitesimal
variations of the maps $Z$ of the form
$$\delta Z=\theta^--2\theta^0Z+\theta^+Z^2, \eqno(3.3)$$
where
$$\pmatrix{\hphantom{-}\theta^0{}_\beta&\hphantom{-}\theta^-{}_\beta\cr
-\theta^+{}_\beta&-\theta^0{}_\beta\cr}
\pmatrix{a_{\beta\alpha}&b_{\beta\alpha}\cr
c_{\beta\alpha}&d_{\beta\alpha}\cr}=
\pmatrix{a_{\beta\alpha}&b_{\beta\alpha}\cr
c_{\beta\alpha}&d_{\beta\alpha}\cr}
\pmatrix{\hphantom{-}\theta^0{}_\alpha&\hphantom{-}\theta^-{}_\alpha\cr
-\theta^+{}_\alpha&-\theta^0{}_\alpha\cr}.\eqno(3.4)$$
The last condition means geometrically that the matrix functions $\xi=
\Big(\matrix{\hphantom{-}\theta^0&\hphantom{-}\theta^-\cr-\theta^+&-\theta^0\cr}
\Big)$ define an infinitesimal gauge transformation of the bundle $\Phi$.
The corresponding variations for $\rho$ and $\mu$ are expressed in terms of
a single field $\epsilon$ given by
$$\epsilon=\big(\theta^--2\theta^0Z+\theta^+Z^2\big)/\partial Z.\eqno(3.5)$$
As can be shown easily, $\epsilon$ is independent from map choices in
$\sans A$ and transforms as
$$\epsilon_b=k^{-1}{}_{ba}\epsilon_a\eqno(3.6)$$
under coordinate changes in $\sans a$. One has then
$$\delta\rho=\big(-(1/2)\partial^3+2\rho\partial+(\partial\rho)\big)\epsilon,
\eqno(3.7)$$
$$\delta\mu=\big(\bar\partial-\mu\partial+(\partial\mu)\big)\epsilon.
\eqno(3.8)$$
By construction, the variations $(3.7)-(3.8)$ leave  the flat
$SL(2,{\bf C})$ vector bundle $\Phi$ invariant. On the other hand they are
precisely the form of the $W_2$ symmetry transformation which appear in the
literature. Now, from the definitions of generalized projective structure and
the construction of the associated generalized projective connections and
Beltrami differentials given in sect. 2, it should be fairly clear how $W_n$
symmetries emerge in classical light cone $W_n$ geometry. Here, I shall adopt
from the start an infinitesimal point of view. As will be shown in due course,
$W_n$ symmetries will be given as variations of the geometrical fields
expressible in terms of $n-1$ fields $\epsilon_i$.

Let $\sans A$ be a generalized
projective structure and let $\Phi$ be the associated flat $SL(n,{\bf C})$
vector bundle. The infinitesimal variations of the maps $Z$ of $\sans A$ will
be written in terms of a ghost matrix field $\gamma$ transforming under changes
of trivialization of the bundle $\Phi$ as
$$\gamma_\beta\Phi_{\beta\alpha}=\Phi_{\beta\alpha}\gamma_\alpha\eqno(3.9)$$
and having zero trace
$${\rm tr}\gamma=0.\eqno(3.10)$$
The matrix elements of $\gamma$ are nilpotent, i. e.
$$(\gamma^r{}_s)^2=0,\quad r,s=0,\cdots,n-1.\eqno(3.11)$$
The operation of infinitesimal variation is given by the action of the
Slavnov operator $s$, which is nilpotent:
$$s^2=0.\eqno(3.12)$$
By generalizing the relations $(3.3)$ to the present context, one finds
$$sZ^r=\gamma^r{}_sZ^s-\gamma^0{}_sZ^rZ^s,\eqno(3.13)$$
$$s\gamma=\gamma^2.\eqno(3.14)$$
As may be checked, the nilpotency relation $(3.12)$ is fulfilled. In the
following treatment, the relevant combination of ghost fields is given by
$$C=\sum_{k=0}^{n-1}\kappa^{(k)}W\partial^k\hat\gamma^t W^{-1},\eqno(3.15)$$
where
$$\hat\gamma=\gamma-{1\over n}\sum_{k=0}^{n-1}{\rm tr}
\big(\kappa^{(k)}W\partial^k\gamma^t W^{-1}\big)1,\eqno(3.16)$$
with $\kappa^{(k)}{}_i{}^j=\Big({i\atop k}\Big)\delta_{i-k,j}$ and $W$ defined
by $(2.19)$. The ghost field $C$ is independent from map choices in $\sans A$.
Under coordinate changes in $\sans a$, $C$ transforms as
$$C_b=\Lambda_{ba}C_a\Lambda_{ba}{}^{-1}.\eqno(3.17)$$
Further, $C$ is traceless
$${\rm tr}C=0.\eqno(3.18)$$
In terms of the ghost field $C$, one has that
$$sW=CW,\eqno(3.19)$$
$$s\Omega=\partial C+[C,\Omega],\eqno(3.20a)$$
$$s\Omega^*=\bar\partial C+[C,\Omega^*],\eqno(3.20b)$$
$$sC=C^2.\eqno(3.21)$$
As can be easily checked, the identities $(3.19)-(3.21)$ are consistent
with the nilpotency relation $(3.12)$.

\noindent{\it Proof}. Let $\Xi$ be a smooth $SL(n,{\bf C})$--valued local
function. Set
$$Z'{}^r={\Xi^r{}_sZ^s\over\Xi^0{}_tZ^t}.\eqno(3.22)$$
Then, it is straightforward to check that
$$\Delta'=\Delta{{\rm \det}\Big(\sum_{k=0}^{n-1}\kappa^{(k)}\tilde W
\partial^k\Xi^t\tilde W^{-1}\Big)\over\big(\Xi^0{}_tZ^t\big)^n},\eqno(3.23)$$
where $\Delta'$ is defined in terms of the $Z'{}^r$'s according to eqs.
$(2.9)-(2.10)$ with the $Z^r$'s replaced by the $Z'{}^r$'s and the
numerical matrix $\kappa^{(k)}$ is defined below eq. $(3.16)$. From $(2.19)$,
it is simple to show that
$$W=\Psi\tilde W\eqno(3.24)$$
with
$$\eqalignno{\Psi_i{}^j&={i\choose j}\partial^{i-j}\Delta^{-{1\over n}},
\quad i\geq j,&(3.25)\cr\Psi_i{}^j&=0,\quad i<j.&\cr}$$
One can further check that
$$[\Psi,\kappa^{(k)}]=0. \eqno(3.26)$$
By using $(3.24)-(3.26)$, it is readily seen that one can replace $\tilde W$
by $W$ in the numerator of the fraction in the right hand side of $(3.23)$.
Now, set $\Xi=1+\xi$ with ${\rm tr}\xi=0$ and linearize the right hand side
of eq. $(3.23)$ with respect to $\xi$. By doing so, one finds easily that
$$s\Delta=-n\Big[\gamma^0{}_sZ^s-{1\over n}\sum_{k=0}^{n-1}{\rm tr}
\big(\kappa^{(k)}W\partial^k\gamma^t W^{-1}\big)\Big]\Delta.\eqno(3.27)$$
{}From this relation and $(3.13)$, it follows that
$$s\big(\Delta^{-{1\over n}}Z^r\big)=\hat\gamma^r{}_s\Delta^{-{1\over n}}Z^s,
\eqno(3.28)$$
where $\hat\gamma$ is defined by eq. $(3.16)$. From $(2.19)$, it is then
straightforward to deduce $(3.19)$ with $C$ given by $(3.15)-(3.16)$.
{}From $(2.21)$, $(3.9)$ and $(3.15)$, it follows that the ghost matrix
field $C$ is independent from map choices in $\sans A$. It is clear that the
Slavnov operator $s$ commutes with coordinate changes in $\sans a$. From this
remark and $(2.21)$, it follows that under coordinate changes in $\sans a$,
$C$ transforms as in $(3.17)$. The tracelessness relation $(3.18)$ is
straightforward to verify from $(3.15)-(3.16)$. $(3.20a)-(3.20b)$ follow
readily from $(2.25a)-(2.25b)$ and $(3.19)$. There remain to verify only
$(3.21)$. To begin with, one notes that
$$\hat\gamma^2=\gamma^2\eqno(3.29)$$
because of the anticommuting character of the matrix elements of the ghost
field $\gamma$ (cfr. eq. $(3.11)$).
{}From $(3.14)$, $(3.15)$, $(3.19)$ and $(3.29)$, one verifies
straightforwardly that
$$s\sum_{k=0}^{n-1}\kappa^{(k)}W\partial^k\gamma^t W^{-1}=C^2.\eqno(3.30)$$
By using $(3.14)$, $(3.29)$ and $(3.30)$, one finds that
$$s\hat\gamma=\gamma^2-(1/n){\rm tr}\big(C^2\big)=\gamma^2=\hat\gamma^2.
\eqno(3.31)$$
Then, from $(3.15)-(3.16)$, through a calculation similar to that leading
to $(3.30)$, one deduces straightforwardly $(3.21)$. {\it QED}.

The matrix elements of the ghost field $C$ are not all independent. In fact,
they can all be expressed in terms of $n-1$ elements $\epsilon_i$ given by
$$\epsilon_i=C_i{}^{n-1},\quad i\leq n-2\eqno(3.32)$$
and the $\rho^i$'s. The explicit expression of the generic matrix element
$C_i{}^j$ of $C$ is
$$\eqalignno{C_i{}^j&=-\sum_{k=1}^{n-2-j}\sum_{l=0}^{n-2-j-k}(-1)^l
{n-2-j-k\choose l}\partial^l\big(\epsilon_{i+n-2-j-k-l}\rho^{n-1-k}\big)
&(3.33a)\cr
&\hskip2.7cm+\sum_{l=0}^{n-1-j}(-1)^l{n-1-j\choose l}\partial^l
\epsilon_{i+n-1-j-l},\quad i<j,&\cr
C_i{}^j&=-\sum_{k=1}^{n-2-i}\sum_{l=0}^{n-2-i-k}(-1)^l{n-2-i-k\choose l}
\partial^l\big(\epsilon_{n-2-k-l}\rho^{n-1-k}\big)&(3.33b)\cr
+&\sum_{l=1}^{n-1-i}(-1)^l{n-1-i\choose l}
\partial^l\epsilon_{n-1-l}
-{1\over n}\Bigg[\sum_{l=1}^{n-1}(-1)^l{n\choose l+1}
\partial^l\epsilon_{n-1-l}&\cr
&\hskip.6cm-\sum_{k=1}^{n-2}\sum_{l=0}^{n-2-k}(-1)^l
{n-1-k\choose l+1}\partial^l\big(\epsilon_{n-2-k-l}\rho^{n-1-k}\big)
\Bigg],\quad i=j,&\cr
C_i{}^j&=-\sum_{k=1}^{n-2}\sum_{l=0}^{n-2-k}(-1)^l\Bigg[
\sum_{m=0}^{\langle n-2-k-l,j\rangle}{i-1-m\choose i-j-1}{n-2-k-m\choose l}
&(3.33c)\cr
&\hskip1cm-{1\over n}{i\choose j}{n-1-k\choose l+1}\Bigg]\partial^{i-j+l}
\big(\epsilon_{n-2-k-l}\rho^{n-1-k}\big)&\cr
&\hskip1cm+\sum_{k=0}^j\sum_{l=0}^{i-j-1}{l+k \choose k}
\partial^l\big(\epsilon_{i-1-k-l}\rho^{j-k}\big)&\cr
+&\sum_{l=1}^{n-1}(-1)^l\Bigg[
\sum_{m=0}^{\langle n-1-l,j\rangle}{i-1-m\choose i-j-1}{n-1-m\choose l}
&\cr
&\hskip3.41cm-{1\over n}{i\choose j}{n\choose l+1}\Bigg]\partial^{i-j+l}
\epsilon_{n-1-l},\quad i>j, &\cr}$$
where the notational conventions stated below eq. $(2.32)$ hold.

\noindent{\it Proof}. From $(2.31a)$ and $(3.20a)$, it is apparent that,
for $i\leq n-2$, $s\Omega_i{}^j=0$ identically. This identity entails a host
of relations between the matrix elements of the ghost field $C$. As is
easily seen, such relations read
$$C_i{}^j-C_{i-1}^{\hphantom{1}j-1}
-\partial C_{i-1}^{\hphantom{1}j}=
C_{i-1}^{\hphantom{1}n-1}\Omega_{n-1}^{\hphantom{1}j},\quad 1\leq i,
\eqno(3.34)$$
where $C_i{}^{-1}=0$ by convention.
To this, one has to add the tracelessness relation $(3.18)$. As
appears, these relations and constraints are formally identical to $(2.33)$
with $\Omega^*$ replaced by $C$. So, proceeding as in the derivation $(2.32)$,
one finds that there is a unique solution of those relations once $C_i{}^{n-1}$
is specified. This is obtainable from $(2.32)$ by performing the substitutions
$\Omega^*{}_i{}^j\rightarrow C_i{}^j$ and $\mu_i\rightarrow\epsilon_i$.
{\it QED}.

It is clear that the $\epsilon_i$'s are independent from map choices in
$\sans A$. It can also be checked from $(3.17)$ that
$$\epsilon_{bi}=k^{{1\over 2}(1-n)}{}_{ba}\sum_{j=0}^{n-2}\Lambda_{ba}{}_i{}^j
\epsilon_{aj}
\eqno(3.35)$$
under coordinate changes in $\sans a$.

The expressions of $s\rho^i$, $s\mu_i$ and $s\epsilon_i$ can be obtained from
$(3.20)-(3.21)$ by specializing to the relevant matrix elements. Such
expressions read
$$s\rho^i=\partial C_{n-1}^{\hphantom{1}i}+C_{n-1}^{\hphantom{1}i-1}
+\rho^iC_{n-1}^{\hphantom{1}n-1}-\sum_{k=0}^{n-2}\rho^kC_k{}^i,\eqno(3.36)$$
$$s\mu_i=\bar\partial\epsilon_i+\epsilon_i\Omega^*{}_{n-1}^{\hphantom{1}n-1}
-\Omega^*{}_i{}^{n-1}C_{n-1}^{\phantom{1}n-1}
+\sum_{k=0}^{n-2}\big(C_i{}^k\mu_k-\Omega^*{}_i{}^k\epsilon_k\big),\eqno(3.37)$$
$$s\epsilon_i=\sum_{j=0}^{n-2}\big(C_i{}^j-\delta_i{}^jC_{n-1}^{\hphantom{1}n-1}
\big)\epsilon_j,\eqno(3.38)$$
where the matrix elements $\Omega^*{}_i{}^j$ are given in terms of the
$\rho^i$'s and the $\mu_i$'s by $(2.32)$ and the matrix elements $C_i{}^j$
are given in terms of the $\rho^i$'s and $\epsilon_i$ by $(3.33)$ and
$C_{n-1}^{\hphantom{1}-1}=0$ by convention. $(3.36)-(3.38)$ provide the
complete
BRS algebra of $W_n$ symmetry.

To summarize, I have shown that $W_n$ symmetries correspond to deformations of
the geometric fields $\rho^i$ and $\mu_i$ which do not change the flat
$SL(n,{\bf C})$ vector bundle associated to the underlying generalized
projective structure. It is readily checked that, for $n=2$, $\epsilon_1=
\epsilon$ is given by $(3.5)$ and that $(3.35)-(3.37)$ reduce into $(3.6)
-(3.8)$. For $n=3$, the combinations of ghost fields commonly used in the
literature are $\tilde\epsilon_0=\epsilon_0$ and $\tilde\epsilon_1=
\epsilon_1-(1/2)\partial\epsilon_0$ and that $s\tilde\rho^0$, $s\tilde\rho^1$,
$s\tilde\mu_0$ and $s\tilde\mu_1$ are given by the same expressions worked out
in refs. \ref{27--30}, where $\tilde\rho^0$, $\tilde\rho^1$, $\tilde\mu_0$ and
$\tilde\mu_1$ are defined in the paragraph following that of eq. $(2.45)$.
\vskip.6cm
\centerline{\bf 4. Field Theoretic Models and Anomaly Analysis.}
\vskip.5cm

In the previous two sections, I have illustrated in detail $W_n$ geometry and
its symmetries. The natural question arises about which field theoretic
implications and applications that framework may have. To see this, two basic
remarks are necessary.

\par\noindent $a$) The ghost fields $\epsilon_i$'s defined in $(3.32)$ are
complicated non local functionals of the generalized projective connections
$\rho^i$ and Beltrami differentials $\mu_i$ and the elemental ghost fields
$\gamma_r{}^s$ via $(3.15)-(3.16)$. On the other hand, the BRS algebra
$(3.36)-(3.38)$ is given completely in terms of local expressions in the
$\rho^i$'s, the $\mu_i$'s and the ghost fields $\epsilon_i$ (cfr. eqs. $(2.32)$
and $(3.33)$).

\par\noindent $b$) $s\rho^i$ and $s\epsilon_i$ are expressed solely in terms of
the $\rho^j$'s and the $\epsilon_k$'s (cfr. eq. $(3.33)$).

\par\noindent
Because of these observations, it is conceivable to consider local field
theoretic models having the following features. The geometric background is
specified by $n-1$ generalized projective connections $\rho^i$ transforming
under coordinate changes in $\sans a$ according to $(2.39)$. The symmetry
transformations of the $\rho^i$'s are given by $(3.36)$ in terms of elementary
ghost fields $\epsilon_i$ transforming under coordinate changes in $\sans a$
according to $(3.35)$ with $C$ given by $(3.33)$. The nilpotency of $s$ (cfr.
eq. $(3.12)$) requires then that $(3.38)$ holds. The matter fields glue in
$\sans a$ in such a way to render the local classical Lagrangian a well-defined
density. The symmetry transformations of the matter fields are given by local
expressions in the matter fields themselves and the $\rho^i$'s and
$\epsilon_i$'s and, of course, must be compatible with the nilpotency of $s$.
The local classical action is invariant under the symmetry transformations of
all fields. Since one is dealing with local field theory, one can analyze
the possible anomalies of the quantum theory by well-known techniques based on
the study of the local cohomology of the Slavnov operator $s$. Models such as
these are examples of chiral projective field theories. As explained in the
introduction, the associated induced actions are related to quantum $W_n$
gravity. Below, I shall provide a few basic examples.
Before proceeding further, however, a technical remark is in order. Though the
independent geometric fields and ghost fields are, respectively, the $\rho^i$'s
and the $\epsilon_i$'s, it is far more convenient to operate through the
derived fields $\Omega$ and $C$ given by $(2.31)$ and $(3.33)$, respectively.
It is clear that $\Omega$ and $C$ glue under coordinate changes in $\sans a$
as in $(2.26a)$ and $(3.17)$, respectively, since their gluing properties are
determined by those of the $\rho^i$'s and the $\epsilon_i$'s. For similar
reasons, the symmetry transformations of $\Omega$ and the $C$ are given by
$(3.20a)$ and $(3.21)$, respectively. Correspondingly, the matter fields
transform under coordinate changes in $\sans a$ according to representations
of the jet bundle $\Lambda$ and have symmetry transformations given locally
in terms of $C$ and the fields themselves. Such formalism is
highly redundant, but has the crucial advantage of producing compact formulae
and rendering manageable otherwise intractable analyses and calculations.

The first model is defined as follows. The matter fields $\psi^*$ and $\phi$
are given locally as smooth complex $n$ vector functions transforming under
coordinate changes in $\sans a$ as
$$\psi^*{}_b=\bar k_{ba}\psi^*{}_a\Lambda_{ab},\eqno(4.1)$$
$$\phi_b=\Lambda_{ba}\phi_a.\eqno(4.2)$$
The local classical action is
$$S(\psi^*,\phi;\rho)=\int_\Sigma{d\bar z\wedge dz\over 2i}
\psi^*\cdot(\partial-\Omega)\phi\eqno(4.3)$$
with $\Omega$ given by $(2.31)$. It is easily
checked from $(2.26)$ and $(4.1)-(4.2)$ that the integrand is a well defined
density so that the right hand side of $(4.3)$ is well-defined. The action
of the Slavnov operator $s$ on $\psi^*$ and $\phi$ is defined by
$$s\psi^*=-\psi^*C,\eqno(4.4)$$
$$s\phi=C\phi.\eqno(4.5)$$
$(3.36)$, $(3.38)$, $(4.4)$ and $(4.5)$ are compatible with $(3.12)$ and
constitute the BRS algebra of the model. It is a straightforward exercise
to verify that
$$sS(\psi^*,\phi;\rho)=0, \eqno(4.6)$$
i. e. $S(\psi^*,\phi;\rho)$ is invariant under the $W_n$ symmetry in the
sense defined earlier in this section. The classical equations of motion are
$$(\partial+\Omega^t)\psi^{*t}=0,\eqno(4.7)$$
$$(\partial-\Omega)\phi=0.\eqno(4.8)$$
By writing these equations in components, one finds that $\psi^*{}_i=
-\partial\psi^{*i+1}-\rho^{i+1}\psi^{*n-1}$ for $i\leq n-2$ with $\rho^{n-1}=0$
and $\phi_j=\partial\phi_{j-1}$ for $1\leq j$, so that
the fields $\psi^*{}^i$ with $i\leq n-2$
and $\phi_j$ with $1\leq j$ are auxiliary. The
dynamical fields $\psi^{*n-1}$ and $\phi_0$ obey the equations
$$\bigg((-1)^n\partial^n-\sum_{i=0}^{n-2}(-1)^i\partial^i\rho^i\bigg)
\psi^{*n-1}=0, \eqno(4.9)$$
$$\bigg(\partial^n-\sum_{j=0}^{n-2}\rho^j\partial^j\bigg)\phi_0=0.
\eqno(4.10)$$
Note that the differential operators showing here correspond to the $n$--th
reduction of the KP hierarchy.

Another local field theoretic model is the following. The basic fields
$\Psi^*$ and $\Phi$ are given locally as smooth $n\times n$ matrix functions
transforming under coordinate changes in $\sans a$ as
$$\Psi^*{}_b=\bar k_{ba}\Lambda_{ba}\Psi^*{}_a\Lambda_{ba}{}^{-1},\eqno(4.11)$$
$$\Phi_b=\Lambda_{ba}\Phi_a\Lambda_{ba}{}^{-1}.\eqno(4.12)$$
It is further assumed that
$${\rm tr}\Psi^*=0,\eqno(4.13)$$
$${\rm tr}\Phi=0.\eqno(4.14)$$
The local classical action is
$$S(\Psi^*,\Phi;\rho)=\int_\Sigma{d\bar z\wedge dz\over 2i}{\rm tr}\Big[
\Psi^*(\partial\Phi+[\Phi,\Omega])\Big],\eqno(4.15)$$
with $\Omega$ given by $(2.31)$. From $(2.26)$ and $(4.11)-(4.12)$ it follows
that the integrand is a well defined density so that the right hand side of
$(4.15)$ is well-defined. The action of the Slavnov operator $s$ on $\Psi^*$
and $\Phi$ is defined by
$$s\Psi^*=[C,\Psi^*],\eqno(4.16)$$
$$s\Phi=[C,\Phi].\eqno(4.17)$$
As in the case of the previous example, $(3.36)$, $(3.38)$, $(4.16)$ and
$(4.17)$ are compatible with $(3.12)$ and define the BRS algebra of the
model. By means of a straightforward calculation, one verifies that
$$sS(\Psi^*,\Phi;\rho)=0, \eqno(4.18)$$
i. e. $S(\psi^*,\phi;\rho)$ is invariant under the $W_n$ symmetry in the sense
defined earlier in this section. The classical equations of motion are
$$\partial\Psi^*+[\Psi^*,\Omega]=0,\eqno(4.19)$$
$$\partial\Phi+[\Phi,\Omega]=0.\eqno(4.20)$$
By writing these equations in components, one finds that the only matrix
elements of the fields $\Psi^*$ and $\Phi$ which are not auxiliary fields are
$\psi^*{}_i=\Psi^*{}_i^{\hphantom{1]}n-1}$ and
$\phi_i=\Phi_i^{\hphantom{1]}n-1}$ with $i\leq n-2$, respectively.
The matrix elements $\Psi^*{}_i{}^j$ and $\Phi_i{}^j$ are given
in terms of the $\psi^*{}_i$'s and $\phi_i$'s and the $\rho^i$'s by expressions
formally identical to $(3.33)$ with $C_i{}^j$, $\epsilon_k$ replaced by
$\Psi^*{}_i{}^j$, $\psi^*{}_k$ and $\Phi_i{}^j$, $\phi_k$ respectively. The
proof of this statement is analogous to that of $(3.33)$. The
equations of motion for the dynamical components $\psi^*{}_k$ and $\phi_k$
read in concise form
$$\partial\Psi^*{}_{n-1}^{\hphantom{1}i}+\Psi^*{}_{n-1}^{\hphantom{1}i-1}
+\Psi^*{}_{n-1}^{\hphantom{1}n-1}\rho^i-\sum_{j=0}^{n-2}\rho^j\Psi^*{}_j{}^i=0,
\eqno(4.21)$$
$$\partial\Phi_{n-1}^{\hphantom{1}i}+\Phi_{n-1}^{\hphantom{1}i-1}
+\Phi_{n-1}^{\hphantom{1}n-1}\rho^i-\sum_{j=0}^{n-2}\rho^j\Phi_j{}^i=0,
\eqno(4.22)$$
where the matrix elements of $\Psi^*$ and $\Phi$ are given in terms of the
$\psi^*{}_k$'s and the $\phi_k$'s as indicated above. The above examples are
canonical, being based on the defining and adjoint representations of $SL(n,
{\bf C})$. Of course, examples employing other representations of $SL(n,
{\bf C})$ could be easily worked out.

Let us now discuss the quantization of a chiral projective field theory. I
shall leave aside the highly non trivial problem of specifying the quantization
procedure the solution of which is likely to require a covariant formulation
and a holomorphic factorization theorem which are not available at present.
Instead, I shall concentrate on analyzing the following properties of the
induced projective action $\Gamma_n(\rho)$ which are expected to hold from
first principles.

\par\noindent $a$) $\Gamma_n(\rho)$ is a non local holomorphic functional
of the $\rho^i$'s. This follows from chirality.

\par\noindent $b$) $\Gamma_n(\rho)$ obeys a Ward identity of the form
$$s\Gamma_n(\rho)=\kappa{\cal A}_n(\epsilon,\rho), \eqno(4.23)$$
where ${\cal A}_n(\epsilon,\rho)$ is an anomaly and $\kappa$ is a
constant measuring its strength and depending on the matter content of
the model considered. Here, I shall assume that $\kappa$ is non zero.
A complete classification of the possible
anomalies arising in the quantum theory reduces to that of all local
functionals ${\cal A}_n(\epsilon,\rho)$ with ghost number $+1$ satisfying the
Wess-Zumino consistency condition $s{\cal A}_n(\epsilon,\rho)=0$ and such
that ${\cal A}_n(\epsilon,\rho)\not=s{\cal F}(\rho)$ for any local
functional ${\cal F}(\rho)$ of ghost number $0$.

\par\noindent In this paper, I shall limit myself to write down a
functional ${\cal A}_n(\epsilon,\rho)$ which is certain to be the relevant
anomaly since it leads to a Ward identity particular cases of which have
been obtained by alternative approaches for $n=2,3$ \ref{27--30}.
Explicitly, ${\cal A}_n(\epsilon,\rho)$ is given by
$${\cal A}_n(\epsilon,\rho)
=\int_\Sigma{d\bar z\wedge dz\over 2i}{\rm tr}(C\bar\partial\Omega).
\eqno(4.24)$$
{}From $(2.26a)$ and $(3.17)$ and the $\sans a$--holomorphy of the jet bundle
$\Lambda$, it is easily seen that the integrand in $(4.24)$ is a well-defined
density so that the integration makes sense. Let us verify that
${\cal A}_n(\epsilon,\rho)$ fulfills the Wess-Zumino consistency relation.
To the end of performing crucial integration by parts, it is necessary to
introduce a $\sans a$--holomorphic background connection $\Omega^{(0)}$ of the
jet bundle $\Lambda$. Under coordinate changes in $\sans a$, $\Omega^{(0)}$
transforms as $(2.26a)$ with $\Omega$ replaced by $\Omega^{(0)}$ and, further,
$\bar\partial\Omega^{(0)}=0$. The existence of $\Omega^{(0)}$ is guaranteed
by the $\sans a$--holomorphy and flatness of $\Lambda$. Through a simple
calculation based on $(3.20a)$, $(3.21)$, one finds
$$\eqalignno{s{\cal A}_n(\epsilon,\rho)&=
\int_\Sigma{d\bar z\wedge dz\over 2i}{\rm tr}\Big[-C\bar\partial\partial C
+\bar\partial(C\Omega^{(0)}C)++\bar\partial\big(C(\Omega-\Omega^{(0)})C\big)
\Big]&(4.25)\cr
&=\int_\Sigma{d\bar z\wedge dz\over 2i}{\rm tr}\Big[\big(\partial C
+[C,\Omega^{(0)}]\big)\bar\partial C\Big]&\cr
&=0\vphantom{\int_\Sigma{d\bar z\wedge dz\over 2i}}.&\cr}$$
The vanishing of the middle term is proven by performing an integration by
parts and exploiting the $\sans a$--holomorphy of $\Omega^{(0)}$ and the
anticommutativity of the ghost matrix field $C$.
A detailed proof of the non triviality of ${\cal A}_n(\epsilon,\rho)$ is
probably very difficult and will be left to future work. The anomaly
${\cal A}_n(\epsilon,\rho)$ is formally identical to that of ordinary gauge
theory. There is though a crucial difference: the matrix
elements of the connection $\Omega$ and the ghost $C$ are not all independent,
being functionals of the elementary fields $\rho^i$ and $\epsilon_i$ via
$(2.31)$ and $(3.33)$. ${\cal A}_n(\epsilon,\rho)$ has a simple expression
in terms of these latter fields:
$${\cal A}_n(\epsilon,\rho)
=\int_\Sigma{d\bar z\wedge dz\over 2i}
\sum_{i=0}^{n-2}\epsilon_i\bar\partial\rho^i. \eqno(4.26)$$
The identity of $(4.24)$ and $(4.26)$ follows easily from $(2.31)$
and $(3.32)$.

The currents associated to the induced projective action $\Gamma_n(\rho)$
are defined by
$$\mu_i(\rho)=-{1\over\kappa}{\delta\Gamma_n(\rho)\over\delta\rho^i},
\quad i\leq n-2.\eqno(4.27)$$
The $\mu_i(\rho)$'s are non local holomorphic functionals of the $\rho^i$'s.
Under coordinate changes in $\sans a$, the $\mu_i(\rho)$'s transform as
in $(2.40)$. This follows readily from $(2.39)$ and the fact that
$\sum_{i=0}^{n-2}\delta\rho^i\mu_i(\rho)$ is a well-defined density. By means
of the $\mu_i(\rho)$'s one can construct a matrix field $\Omega^*(\rho)$ by
using the formulae $(2.32)$ with the $\mu_i$'s replaced by $\mu_i(\rho)$'s.
It is clear that under coordinate changes in $\sans a$, $\Omega^*(\rho)$
transforms as in $(2.26b)$. Thus, $\Omega$ and $\Omega^*(\rho)$ are the
components of a connection of the jet bundle $\Lambda$. Let us compute the
curvature of this connection. From $(2.29)$, $(2.30)$ and $(2.31a)$, one
obtains easily the variational relation
$$\delta\Gamma_n(\rho)=\int_\Sigma{d\bar z\wedge dz\over 2i}
\sum_{i=0}^{n-2}\delta\rho^i{\delta\Gamma_n(\rho)\over\delta\rho^i}
=\kappa\int_\Sigma{d\bar z\wedge dz\over 2i}
{\rm tr}\big(-\delta\Omega\Omega^*(\rho)\big).\eqno(4.28)$$
{}From $(4.28)$ and $(3.20a)$, one gets then
$$\eqalignno{s\Gamma_n(\rho)&=\vphantom{\sum_{i=0}^{n-2}}\kappa
\int_\Sigma{d\bar z\wedge dz\over 2i}{\rm tr}\big(-s\Omega\Omega^*(\rho)\big)
&(4.29)\cr
&=\vphantom{\sum_{i=0}^{n-2}}\kappa
\int_\Sigma{d\bar z\wedge dz\over 2i}{\rm tr}\Big[C\big(
\partial\Omega^*(\rho)-[\Omega,\Omega^*(\rho)]\big)\Big].&\cr}$$
By combining $(4.29)$ with $(4.23)$ and $(4.24)$, one finds that
$$0=\int_\Sigma{d\bar z\wedge dz\over 2i}{\rm tr}\Big[
C\big(\bar\partial\Omega-\partial\Omega^*(\rho)+[\Omega,\Omega^*(\rho)]\big)
\Big].\eqno(4.30)$$
Now, because of the way $\Omega^*(\rho)$ has been defined above,
$\Omega^*(\rho)$ is traceless and satisfies $(2.33)$ with $\Omega^*$ replaced
by $\Omega^*(\rho)$. By $(2.31a)$, the resulting identities can be cast in
the form $(\bar\partial\Omega-\partial\Omega^*(\rho)+[\Omega,
\Omega^*(\rho)])_i{}^j=0$ for either $i\leq n-2$ or $i=j=n-1$. Thus, the
integrand
in $(4.30)$ reduces into $\sum_{i=0}^{n-2}\epsilon_i(\bar\partial\Omega
-\partial\Omega^*(\rho)+[\Omega,\Omega^*(\rho)])_{n-1}^{\hphantom{1}i}$,
by $(3.32)$. Since the ghost fields $\epsilon_i$ are independent, it follows
from $(4.30)$ that $(\bar\partial\Omega-\partial\Omega^*(\rho)+[\Omega,
\Omega^*(\rho)])_{n-1}^{\hphantom{1}i}=0$ for $i \leq n-2$. Thus, $\Omega$ and
$\Omega^*(\rho)$ fulfil $(2.28)$. This shows that the Ward identity $(4.23)$
is equivalent to the flatness condition of the connection $(\Omega,
\Omega^*(\rho))$ (cfr. eq. $(2.28)$), or, what is the same, to the
compatibility of the $\rho^i$'s and the $\mu_i(\rho)$'s (cfr. $(2.41)$).
The relation between Ward identities and flatness or compatibility conditions
is recurrent in $W_n$ gravity \ref{21--26}.

Let us compute the action of the Slavnov operator on the currents
$\mu_i(\rho)$.
Recall that the operator $\delta$ of differentiation with respect to the
$\rho^i$'s anticommutes with $s$ and the ghost fields $\epsilon_k$.
{}From $(4.23)$ and $(4.24)$, one has that
$$\delta s\Gamma_n(\rho)=\kappa\int_\Sigma{d\bar z\wedge dz\over 2i}
{\rm tr}\Big(\delta C\bar\partial\Omega-\delta\Omega\bar\partial C\Big).
\eqno(4.31)$$
{}From $(4.28)$ and $(3.20a)$, one finds that
$$\eqalignno{s\delta\Gamma_n(\rho)&=\kappa\int_\Sigma{d\bar z\wedge dz\over 2i}
{\rm tr}\Big[\delta s\Omega\Omega^*(\rho)+\delta\Omega s\Omega^*(\rho)
\Big]&(4.32)\cr
&=\kappa\int_\Sigma{d\bar z\wedge dz\over 2i}
{\rm tr}\Big[\delta\Omega\big(s\Omega^*(\rho)-[C,\Omega^*(\rho)]\big)
-\delta C\big(\partial\Omega^*(\rho)-[\Omega,\Omega^*(\rho)]\big)\Big].&\cr}$$
By adding $(4.31)$ and $(4.32)$ side by side and using $(2.28)$ and the
relation $\delta s+s\delta=0$, one finds then that
$$0=\kappa\int_\Sigma{d\bar z\wedge dz\over 2i}
{\rm tr}\Big[\delta\Omega\big(s\Omega^*(\rho)-\bar\partial C-[C,\Omega^*(\rho)]
\big)\Big].\eqno(4.33)$$
By $(2.31)$, only $\delta\Omega_{n-1}^{\hphantom{1}i}=\delta\rho^i$ with
$i\leq n-2$ is non vanishing. Thus, the integrand in $(4.33)$ is
$\sum_{i=0}^{n-2}\delta\rho^i\big(s\Omega^*(\rho)-\bar\partial C-
[C,\Omega^*(\rho)]\big)_i^{\hphantom{1}n-1}$. From $(2.32)$ and $(3.33)$,
it follows then that $s\mu_i(\rho)$ is given by $(3.37)$.
This implies further that $s\Omega^*(\rho)$ is given by $(3.20b)$, as
$(3.36)-(3.37)$ are equivalent to $(3.20)$.

As indicated in the introduction, the induced $W_n$ gravity action
$\Gamma^*_n(\mu)$ is related to the generating functional $W_n(\mu)$ of the
connected Green functions of the projective connections $\rho^i$ quantized via
the induced projective action $\Gamma_n(\rho)$. The classical contribution
$W_{n0}(\mu)$ of $W_n(\mu)$, which is the leading term of $W_n(\mu)$ in the
limit $\kappa\rightarrow\infty$, is the functional Legendre transform
of $\Gamma_n(\rho)$. To compute $W_{n0}(\mu)$, one has to invert the
functionals $\mu_i(\rho)$ and regard the generalized Beltrami differentials
$\mu_i$'s as independent geometric fields. This yields $n-1$ generalized
projective connections $\rho^i(\mu)$ depending
holomorphically and non locally on the $\mu_i$'s. By means of the $\mu_i$'s
and the $\rho^i(\mu)$'s, one can build matrix fields $\Omega(\mu)$ and
$\Omega^*(\mu)$ by means of $(2.31)-(2.32)$ with the $\rho^i$'s replaced by
the $\rho^i(\mu)$'s. The geometric fields $\mu_i$ and $\rho^i(\mu)$
satisfy $(2.41)$ or, equivalently, $(2.28)$. As to symmetries, one can
construct a ghost matrix field $C(\mu)$ by means of the ghost fields
$\epsilon_i$ and the $\rho^i(\mu)$ via $(3.33)$ with the $\rho^i$'s replaced
by the $\rho^i(\mu)$'s. The BRS algebra can be written in the equivalent
forms $(3.36)-(3.38)$ or $(3.20)-(3.21)$. $(3.37)$, $(3.38)$
form a subalgebra of the BRS algebra that is not local in $\mu$. By
contrast, in the projective formulation, where the $\rho^i$'s constitute
the independent fields, the subalgebra $(3.36)$, $(3.38)$ is local.

The classical contribution $W_{n0}(\mu)$ of $W_n(\mu)$ is given by
$$W_{n0}(\mu)=\Gamma(\rho(\mu))
+\kappa\int_\Sigma{d\bar z\wedge dz\over 2i}
{\rm tr}\Big[\big(\Omega(\mu)-\Omega^{(0)}\big)\Omega^*(\mu)\Big],
\eqno(4.34)$$
where $\Omega^{(0)}$ is $\sans a$--holomorphic connection of the jet
bundle $\Lambda$ (see the remarks above eq. $(4.25)$).
The introduction of $\Omega^{(0)}$ is necessary to
ensure that the integrand is a density. The $\sans a$--holomorphy of
$\Omega^{(0)}$ is just a simplifying assumption. $\Omega^{(0)}$ is
invariant under symmetry transformations, i. e.
$$s\Omega^{(0)}=0.\eqno(4.35)$$
{}From the Ward identity $(4.23)-(4.24)$, the BRS algebra $(3.20)-(3.21)$ or
$(3.36)-(3.38)$ and $(4.35)$, one finds that
$$sW_{n0}(\mu)=\kappa{\cal A}^*_{n0}(\epsilon,\mu),\eqno(4.36)$$
where ${\cal A}^*_{n0}(\epsilon,\mu)$ is an anomaly given by
$${\cal A}^*_{n0}(\epsilon,\mu)=-\int_\Sigma{d\bar z\wedge dz\over 2i}
{\rm tr}\Big[C(\mu)\big(\partial\Omega^*(\mu)+[\Omega^*(\mu),\Omega^{(0)}]
\big)\Big].\eqno(4.37)$$
The proof of the Ward identity $(4.36)-(4.37)$ follows easily from $(4.23)
-(4.24)$ and $(3.20)$.
Eqs. $(4.34)$ and $(4.37)$ can be written in terms of the elementary fields
$\mu_i$ and $\epsilon_i$ and the functionals $\rho^i(\mu)$. For simplicity, I
shall assume that the background connection $\Omega^{(0)}$ is given by $(2.31)$
in terms of $n-1$ $\sans a$--holomorphic generalized projective connections
$\rho^{(0)i}$. The $\rho^{(0)i}$'s can be chosen to be the coefficients of the
$n$--th Bol operator associated to an ordinary $\sans a$--holomorphic
projective connection $\rho^{(0)}$ (see ref. \ref{40} for explicit formulae).
It can be shown that there is a projective coordinate structure $\sans p$
subordinated to $\sans a$ such that $\rho^{(0)i}=0$ in the coordinates of
$\sans p$. (This is no longer true for generic coordinates of $\sans a$).
The explicit expression of $W_{n0}(\mu)$ reads:
$$W_{n0}(\mu)=\Gamma(\rho(\mu))
+\kappa\int_\Sigma{d\bar z\wedge dz\over 2i}\sum_{i=0}^{n-2}(\rho(\mu)
-\rho^{(0)})^i\mu_i. \eqno(4.38)$$
Working out an explicit formula of the anomaly ${\cal A}^*_{n0}(\epsilon,\mu)$
is more laborious. I shall only sketch the details. Let $\Omega^{*(0)}(\mu)$
and $C^{(0)}$ be the matrix fields obtained from $(2.32)$ and $(3.33)$ with
$\rho^i$ replaced by $\rho^{(0)i}$, respectively. One has the identity
$$\eqalignno{&{\rm tr}\Big[C(\mu)\big(\partial\Omega^*(\mu)+[\Omega^*(\mu),
\Omega^{(0)}]\big)\Big]&(4.39)\cr
=~&{\rm tr}\Big[C(\mu)\big(\partial\Omega^{*(0)}(\mu)+[\Omega^{*(0)}(\mu),
\Omega^{(0)}]\big)\Big]&\cr
+~&{\rm tr}\Big[\big(C(\mu)-C^{(0)}\big)\Big(\partial\big(\Omega^*(\mu)
-\Omega^{*(0)}(\mu)\big)+[\Omega^*(\mu)-\Omega^{*(0)}(\mu),\Omega^{(0)}]
\Big)\Big]&\cr
+~&{\rm tr}\Big[C^{(0)}\Big(\partial\big(\Omega^*(\mu)
-\Omega^{*(0)}(\mu)\big)+[\Omega^*(\mu)-\Omega^{*(0)}(\mu),\Omega^{(0)}]
\Big)\Big].&\cr}$$
{}From $(2.28)$, one has further
$$\eqalignno{&\partial\big(\Omega^*(\mu)
-\Omega^{*(0)}(\mu)\big)+[\Omega^*(\mu)-\Omega^{*(0)}(\mu),\Omega^{(0)}]
&(4.40)\cr
=~&\bar\partial\Omega-\partial\Omega^{*(0)}(\mu)-[\Omega^{*(0)}(\mu),
\Omega^{(0)}]+[\Omega-\Omega^{(0)},\Omega^*(\mu)].&\cr}$$
This identity is to be substituted in $(4.39)$. One has further that
$(\partial\Omega^{*(0)}(\mu)$ $+$ $[\Omega^{*(0)}(\mu),$ $\Omega^{(0)}])_i{}^j$
$=$ $0$ for either $i\leq n-2$ or $i=j=n-1$ since, as explained already in
similar
instances, the parametrization of $\Omega^{*(0)}$ in terms of the $\rho^{(0)i}$
and the $\mu_i$ is the purely algebraic solution of such relation. Similarly,
one has $(\partial C^{(0)}+[C^{(0)},\Omega^{(0)}])_i{}^j=0$ for either
$i\leq n-2$ or $i=j=n-1$. Note finally that $(\Omega^*(\mu)
-\Omega^{*(0)}(\mu))_i^{\hphantom{1}n-1}=0$ for $i\leq n-2$. By using these
facts, one obtains through some straightforward algebra, the following
expression:
$$\eqalignno{{\cal A}^*_{n0}(\epsilon,\mu)
&=-\int_\Sigma{d\bar z\wedge dz\over 2i}\bigg\{\sum_{i=0}^{n-2}\epsilon_i
\Big[\partial\Omega^{*(0)}{}_{n-1}^{\hphantom{1}i}(\mu)
+\Omega^{*(0)}{}_{n-1}^{\hphantom{1}i-1}(\mu)
&(4.41)\cr
&\hphantom{=-\int_\Sigma{d\bar z\wedge dz\over 2i}\bigg\{\sum_{i=0}^{n-2}
\epsilon_i}
+\Omega^{*(0)}{}_{n-1}^{\hphantom{1}n-1}(\mu)
\rho^{(0)i}-\sum_{j=0}^{n-2}\Omega^{*(0)}{}_j{}^i(\mu)\rho^{(0)j}\Big]&\cr
&\hphantom{=}
-\sum_{i,j=1}^{n-2}\Big[(C(\mu)-C^{(0)})_i{}^j-\delta_i{}^j(C(\mu)
-C^{(0)})_{n-1}^{\hphantom{1}n-1}\Big]\mu_j\big(\rho(\mu)-\rho^{(0)}
\big)\vphantom{)}^i\bigg\}.&\cr}$$
As can be checked easily from $(2.32)$, for $n=2$ this expression gives the
standard anomaly $\sim\int\epsilon_0\partial^3\mu_0$ in coordinates of
$\sans p$. It is slightly harder to verify that for $n=3$, the anomaly can be
written in the standard form $\sim\int\tilde\epsilon_1\partial^3\tilde\mu_1
-(1/12)\tilde\epsilon_0\partial^5\tilde\mu_0$ in coordinates of $\sans p$
\ref{27--28}, where the tilded fields $\tilde\mu_i$ and $\tilde\epsilon_i$
have been defined in sects. 2 and 3, respectively.

To conclude, one notes that the $\rho^i(\mu)$'s are the currents associated to
$W_{n0}(\mu)$. In fact
$$\rho^i(\mu)=\rho^{(0)i}+{1\over\kappa}{\delta W_{n0}(\mu)\over
\delta\mu_i}.\eqno(4.42)$$
This represents the tree level classical contribution the currents of the
$W_n$ algebra.

One could also compute one loop corrections to $W_{n0}(\mu)$ by means of
standard functional methods. This leads to an implicit expression in terms of
the second functional derivative of the classical action $\Gamma_n(\rho)$.
\vskip.6cm
\centerline{{\bf 5. Toward a Covariant} ${\bf W}_{\bf n}$ {\bf Geometry.
Conclusions.}}
\vskip.5cm
Presently, the only known rigorous method of computing the partition
function of a chiral model relies on the holomorphic factorization of the
partition function of the corresponding covariant model \ref{31--32}.
In order to put the formulation of chiral projective field theory expounded
in the previous section on a sounder basis, a covariant formulation would be
desirable. To this end, it is necessary to provide the space of matter
fields with a Hilbert inner product the definition of which requires the
introduction of an arbitrary Hermitian metric $\eta$ on the jet bundle
$\Lambda$. For instance, in the case of the chiral projective field theory
defined in $(4.1)-(4.3)$, the inner products would be $\langle\psi^*,\psi^{*
\prime}\rangle$ $=$ $\int\psi^{*\bar\imath}\psi^{*\prime j}\eta_{\bar\imath j}$
and $\langle\phi,\phi'\rangle$ $=$ $\int\phi_{\bar\imath}
\phi'_j\eta^{\bar\imath j}\eta_{\bar 1 1}{}^{2\over 1-n}$. From
the Hilbert space structure of matter field space, one can then compute the
Hilbert adjoint of the relevant differential operator $\cal D$ appearing in the
classical action. ${\cal D}=\partial-\Omega$ in the example quoted. Presumably,
the dependence of the determinant ${\rm det}{\cal D}^\dagger{\cal D}$ on the
metric $\eta$ can be extracted by using heat kernel methods and a holomorphic
factorization theorem can be shown yielding the chiral determinant ${\rm det}
{\cal D}$. The main technical obstacle to carrying out
such analysis is that nothing is known about the zero mode structure of the
classical theory, the $W_n$ generalization of the Riemann-Roch theory
being not known at present.

Another approach to a covariant formulation of $W_n$ gravity may invoke the
extrinsic geometry of regular non singular embeddings of $\Sigma$ into
the manifolds ${\bf W}_n$ discussed at the end of sect. 3.
It is not difficult to see that the standard hyperbolic Kaehler
metric $g_{r\bar s}=2\partial_{\zeta_r}\bar\partial_{\zeta_s}\ln\big(1-
|\zeta|^2\big)$ of $B_{n-1}({\bf C})$ induces a Kaehler metric on
${\bf W}_n$ with constant negative holomorphic sectional curvature.
For $n=2$, this reproduces the usual geometric setting employed in the
covariant formulation of conformal field theory. For $n>2$, this suggests
an approach on the same lines as that of ref. \ref{26} whose connection
with Toda field theory is ascertained. However, its implementation
remains a challenging problem.
\vskip.6cm
\par\noindent
{\bf Acknowledgements}. I wish to voice my gratitude to V. Ancona,
F. Bastianelli, S. Lazzarini, R. Stora and K. Yoshida for many helpful
discussions.
\vfill\eject
\centerline{\bf REFERENCES}

\def\ref#1{\lbrack #1\rbrack}
\def\NP#1{Nucl.~Phys.~{\bf #1}}
\def\PL#1{Phys.~Lett.~{\bf #1}}

\def\CMP#1{Commun.~Math.~Phys.~{\bf #1}}

\def\MPL#1{Mod.~Phys.~Lett.~{\bf #1}}
\def\IJMP#1{Int.~J.~Mod.~Phys.~{\bf #1}}

\def\AP#1{Ann.~Phys.~{\bf #1}}
\vskip.4cm
\par\noindent

\item{\ref{1}}
A. B. Zamolodchikov, Theor. Math. Phys. {\bf 65} (1985) 1205.

\item{\ref{2}}
V. A. Fateev and A. B. Zamolodchikov, \NP{B280} [{\bf FS18}] (1987) 644.

\item{\ref{3}}
F. A. Bais, P. Bouwknegt, M. Surridge and K. Schoutens, \NP{B304} (1988) 348
and
\NP{B304} (1988) 371.

\item{\ref{4}}
F. A. Bais, T. Tjin and P. van Driel, \NP{B357} (1991) 632.

\item{\ref{5}}
M. Bershadsky and H. Ooguri, \CMP{126} (1989), 49

\item{\ref{6}}
J. Balog, L. Feher, P. Forgacs, L. O'Raifeartaigh and A. Wipf, \PL{B227}
(1989) 214 and \AP{203} (1990) 76.

\item{\ref{7}}
A. Bilal and J.-L. Gervais, \PL{B206} (1988) 412, \NP{B314} (1989)
646 and \NP{B318} (1989) 579.

\item{\ref{8}}
C. Hull, \NP{B364} (1991) 621.

\item{\ref{9}}
K. Yamagishy, \PL{B205} (1988) 466.

\item{\ref{10}}
P. Mathieu, \PL{B208} (1988) 101

\item{\ref{11}}
I. Bakas, \PL{B213} (1988) 313, \NP{B302} (1988) 189 and \CMP{123}
(1989) 627.

\item{\ref{12}}
P. Di Francesco, C. Itzykson and J.-B. Zuber, \CMP{140} (1991) 543.

\item{\ref{13}}
M. Awada and Z. Qiu, \PL{B245} (1990) 359.

\item{\ref{14}}
R. Dijkgraaf, E. Verlinde and H. Verlinde, \NP{B348} (1991) 435

\item{\ref{15}}
J. Goeree, \NP{B358} (1991) 737.

\item{\ref{16}}
A. M. Polyakov, \MPL{A2} no. 11 (1987), 893 and \IJMP{A5} (1990) 833.

\item{\ref{17}}
V. G. Knizhnik, A. M. Polyakov and A. B. Zamolodchikov \MPL{A3} (1988) 819.

\item{\ref{18}}
K. Schoutens. A. Sevrin and P. van Nieuwenhuizen, \PL{B243} (1990) 245,
\PL{B251} (1990) 355 and \NP{B349} (1991) 791.

\item{\ref{19}}
E. Bergshoeff, C. N. Pope and K. S. Stelle, \PL{B249} (1990) 208.

\item{\ref{20}}
C. M. Hull, \PL{B269} (1991) 257.

\item{\ref{21}}
A. Gerasimov, A. Levin and A. Marshakov, \NP{B360} (1991) 537.

\item{\ref{22}}
A. Bilal, V. V. Fock and I. I. Kogan, \NP{B359} (1991) 635.

\item{\ref{23}}
K.Yoshida, ROME--818--1991.

\item{\ref{24}}
G. Sotkov, M. Stashnikov and C. J. Zhu, \NP{356} (1991) 245.

\item{\ref{25}}
G. Sotkov and M. Stashnikov, \NP{B356} (1991) 439.

\item{\ref{26}}
J.-L. Gervais and Y. Matsuo, \PL{B274} (1992) 309 and LEPTENS--91/35.

\item{\ref{27}}
H. Ooguri, K. Schoutens. A. Sevrin and P. van Nieuwenhuizen,
Commun. Math. Phys. {\bf 145} (1992) 515.

\item{\ref{28}}
K. Schoutens. A. Sevrin and P. van Nieuwenhuizen, \NP{B364} (1991) 584,
ITP--SB--91--21, ITP--SB--91--50 and ITP--SB--91--54.

\item{\ref{29}}
C. M. Hull, \PL{B265} (1991) 347, \NP{B367} (1991) 731 and QMW/
PH/91/14.

\item{\ref{30}}
C. M. Hull and L. Palacios, \MPL{A6} (1991) 2995.

\item{\ref{31}}
A. A. Belavin and V. G. Knizhnik, \PL{B168} (1986), 201.

\item{\ref{32}}
M. Knecht, S. Lazzarini and R. Stora, \PL{B262} (1991) 25 and Phys. Lett.
{\bf B273} (1991) 63.

\item{\ref{33}}
A. Ceresole, M. Frau, J. McCarthy and A. Lerda, \PL{B265} (1991) 72.

\item{\ref{34}}
J. de Boer and J. Goeree, \PL{B274} (1992) 289

\item{\ref{35}}
R. Gunning, {\it Lectures on Riemann surfaces}, Princeton University Press
(1966) and references therein.

\item{\ref{36}}
R. Gunning, {\it Lectures on Vector Bundles on Riemann surfaces}, Princeton
University Press (1967) and references therein.

\item{\ref{37}}
S. Kobayashi and K. Nomizu, {\it Foundations of Differential Geometry},
vols. I and II, J. Wiley \& Sons (1963).

\item{\ref{38}}
S. Nag, {\it The Complex Analytic Theory of Teichmueller Spaces},
Canadian Mathematical Society Series of Monographs and Advanced Texts,
J. Wiley \& Sons (1988) and references therein.

\item{\ref{39}}
R. Gunning, {\it On Uniformization of Complex Manifolds: the Role of
Connections}, Princeton University Press (1978) and references therein.

\item{\ref{40}}
F. Gieres, CERN--TH.5985/91.

\bye